\documentclass[prb,twocolumn,floatfix,showpacs,superscriptaddress]{revtex4-1}
\usepackage{epsfig,subfigure,amssymb,amscd,amsmath,graphicx,amsfonts,mathtools,mathcomp,pbox,amssymb}
\usepackage{times,longtable,math dots,hyperref}

\newcommand{\non}{\nonumber}

\newcommand{\ba}{\begin{align}}
\newcommand{\ea}{\end{align}}

\newcommand{\ket}[1]{     |    \,    #1    \rangle}
\newcommand{\bra}[1]{  \langle #1  \,  |}

\newcommand{\CC}{\mathbb{C}}
\newcommand{\RR}{\mathbb{R}}
\newcommand{\QQ}{\mathbb{Q}}
\newcommand{\ZZ}{\mathbb{Z}}

\newcommand{\Mod}[1]{\ (\text{mod}\ #1)}
\usepackage{tabularx}

\usepackage[usenames]{color}

\makeatletter
\newcommand*\dashline{\rotatebox[origin=c]{90}{$\dabar@\dabar@\dabar@$}}
\makeatother

\begin{document}

\title{Parafermionic clock models and quantum resonance}

\author{N. Moran}
\affiliation{Department of Mathematical Physics, Maynooth University, Ireland.}
\author{D. Pellegrino}
\affiliation{Department of Mathematical Physics, Maynooth University, Ireland.}
\author{J. K. Slingerland}
\affiliation{Department of Mathematical Physics, Maynooth University, Ireland.}
\affiliation{Dublin Institute for Advanced  Studies, School of Theoretical  Physics, 10 Burlington Rd, Dublin, Ireland.}
\affiliation{Rudolf Peierls Centre for Theoretical Physics, 1 Keble Road, Oxford OX1 3NP, United Kingdom.}
\author{G. Kells}
\affiliation{Dublin Institute for Advanced  Studies, School of Theoretical  Physics, 10 Burlington Rd, Dublin, Ireland.}

\begin{abstract}
We explore the $\ZZ_{N}$ parafermionic clock-model generalisations of the p-wave
Majorana wire model. In particular we examine whether zero-mode operators
analogous to Majorana zero-modes can be found in these models when one
introduces chiral parameters to break time reversal symmetry. The existence of
such zero-modes implies $N$-fold degeneracies throughout the energy spectrum. We
address the question directly through these degeneracies by characterising the
entire energy spectrum using perturbation theory and exact diagonalisation. We
find that when $N$ is prime, and the length $L$ of the wire is finite, the
spectrum exhibits degeneracies up to a splitting that decays exponentially with
system size, for generic values of the chiral parameters. However, at particular
parameter values (resonance points), band crossings appear in the unperturbed
spectrum that typically result in a splitting of the degeneracy at finite order.
We find strong evidence that these preclude the existence of strong zero-modes
for generic values of the chiral parameters. In particular we show that in the
thermodynamic limit, the resonance points become dense in the chiral parameter
space. When $N$ is not prime, the situation is qualitatively different, and
degeneracies in the energy spectrum are split at finite order in perturbation
theory for generic parameter values, even when the length of the wire $L$ is
finite. Exceptions to these general findings can occur at special
``anti-resonant'' points. Here the evidence points to the existence of strong
zero modes and, in the case of the achiral point of the the $N=4$ model, we are
able to construct these modes exactly.
\end{abstract}


\date{\today} \maketitle

There has recently been growing interest in a class of $\ZZ_N$ symmetric one
dimensional lattice models known as parafermion chain or quantum clock models
\cite{Fendley2012,Fendley2014, Jermyn2014, Cobanera2014, Alexandradinata2015,
Iemini2016}. These models generalise the Kitaev wire model \cite{Kitaev2001}
which exhibits localised unpaired Majorana zero-modes at each end. The recent
surge of interest is inspired in part by proposals for their physical
realisation and their potential application to universal topological quantum
computation\cite{Clarke2013,Alicea2015,Mong2014}, something which is not
possible with Majorana zero-modes.

Clock-like systems of this type have been studied
earlier\cite{Howes1983,Gehlen1985}, and much is known about the exactly solvable
chiral Potts models that occur at special values of the coupling constants (see
e.g.~Refs~\onlinecite{Au-Yang1997,Perk2015}). From the perspective of
topological quantum computation however, there are a number of interesting open
problems surrounding the topological degeneracies and potential zero-modes of
the models. In particular, one may ask whether zero-modes exist which
result in topological degeneracies throughout the energy spectrum
\cite{Jermyn2014,Alexandradinata2015}. In this context one speaks of strong
vs.~weak zero-modes (see e.g. Ref. \onlinecite{Alicea2016}) and the question is
important because degeneracies that exist at energies above the ground-state
potentially allow for topologically fault-tolerant quantum devices at higher
temperatures\cite{Huse2013}. This area is of course also interesting on
a fundamental level, as it addresses if and when decoupled/free zero-mode
quasi-particles can exist in complex systems that are otherwise interacting.

We study this question here by direct perturbative and numerical calculations of
the energy spectra of the models. This approach can be seen as complementary to
the iterative position space constructions where (see
Ref.~\onlinecite{Fendley2012,JackKemp2017} for example), a strong zero-mode on a chain of
length $L$ is defined as an operator $\Psi$ such that $[H,\Psi]\rightarrow 0$ as
$L\rightarrow\infty$, with finite-size corrections whose expectation values
vanish exponentially with increasing system size. For a strong zero-mode it is
also required that $\Psi^N$ is proportional to the identity operator for some
integer $N$, and that $\Psi$ does not commute with some operator $Q$ which
generates a discrete symmetry, so $[H,Q]=0$ but $[\Psi,Q]\neq 0$.

In the thermodynamic limit this definition implies topological degeneracies
throughout the spectrum, as the action of $\Psi$ cycles through degenerate states with
different eigenvalues with respect to $Q$. Although directly probing the energy spectrum
lacks the spatial resolution of the aforementioned iterative approach, it has a
number of advantages for this type of study. In particular, the ability to
resolve the degeneracy at any energy scale means that it can be straightforward
to identify weaker variants of the zero-mode. For example we will see
that although there is often no strong zero-mode, the topological degeneracy is
still preserved up to energies far above those of the ground state manifold.
Similarly, we find scenarios where the degeneracy is preserved at all
energies, but only between subsets of the discrete symmetry sectors. In this
latter case it becomes straightforward, within these sub-spaces, to formally
define hidden zero-modes that have a lower $\ZZ_N$ symmetry.

For the Kitaev wire, it is well known that the topological degeneracy is present
throughout the spectrum. This model has strong Majorana zero-modes on the edges
which provide each state with even fermion number a (nearly) degenerate partner
at odd fermion number and vice versa. Of course the Kitaev wire is equivalent to
an Ising spin chain through Jordan-Wigner transformation and its spectrum is
exactly solvable.

The behaviour for generic $\ZZ_N$-parafermion models is more complicated and
depends on the chiral angle $\theta$ (see Eq.~\ref{eqn:hamiltonian}). In
Ref.~\onlinecite{Fendley2012}, it was noted that the $\ZZ_3$ model does not have
strong zero-modes at the time reversal invariant

points ($\theta=n\frac{2\pi}{3}$ and $\theta = n\frac{2\pi}{3} +
\frac{\pi}{3},\quad \forall n = 0, 1, 2$). However an iteratively constructed
perturbative expansion for an exact zero-mode operator was given for regimes
where $\theta$ takes generic values (sufficiently far from the time reversal
invariant points). The radius of convergence of this expansion was conjectured
to be dependent on $\theta$.
 
In Ref.~\onlinecite{Jermyn2014}, the stability of zero-modes in the $\ZZ_3$
model was explored further using perturbation theory, diagonalisation of
approximate hamiltonians, and DMRG. It was shown explicitly that the breakdown of
zero-modes at the time reversal invariant points is due
to domain-wall tunnelling processes which lead to power law splitting of the
degeneracy between excited states. Evidence was then shown for the suppression
of these processes for generic values of $\theta$, leading to
restoration of apparent exponential splitting and ``zero-mode revival''.

A central result of this manuscript is to show that there is in fact power law
splitting of the topological degeneracy (and hence no strong zero-modes) in
large regions of the model's parameter space. This can be explained in terms of
perturbative effects occurring at resonant crossings between bands labelled by
different types and numbers of generalised domain-wall excitations. Sufficiently
far from these crossings, the degeneracy may be recovered and this is observed
for finite systems. However as the system size increases, so too does the number
of resonant band crossings, spoiling the degeneracy over ever larger areas of
parameter space. Our analysis suggests that in the limit $L\rightarrow\infty$,
strong zero-modes do not exist for generic values of the chiral parameter
$\theta$, at any value of $N$ (although there are at least isolated
$\theta$-values that allow for strong zero-modes for $N=3$ and
guaranteed exactly solvable points for $N=4$). This also implies that the radius
 of convergence of the expansion for the zero-mode given in
Ref.~\onlinecite{Fendley2012} in $\theta$-space is generically zero.

To show the power law splitting, we use degenerate perturbation methods like
those employed in Ref.~\onlinecite{Kells2015} for the interacting Kitaev chain. In a
similar way we find evidence that the topological degeneracy is broken only at
order $\sim L$, provided that the un-perturbed system does not contain any band
crossings. However, because these crossings shift and become more dense as we
make $L$ longer, we cannot say that strong zero-modes exist in the
$L\rightarrow\infty$ limit. Therefore while topological degeneracies can persist at
finite size, and also for many low lying energy levels in the large $L$ limit,
they do not hold generally at finite energy density. It will be outlined
elsewhere \cite{Kellsinprep} why similar scaling applies to the interacting
Kitaev chain.

We consider the model at any value of $N$ and note that there is an important
difference between the cases where $N$ is prime and those where it is not. In
particular we note that if $N$ is composite, certain bands with different number
and type of domain-walls are degenerate for all values of the chiral parameter.
This leads to a uniform power law splitting of the topological degeneracy and
hence no $\ZZ_N$ parafermionic zero-modes occur for generic $\theta$ when $N$ is
composite. Of course the actual splitting varies with $\theta$ and exceptional
values of $\theta$ where the degeneracy is preserved can occur.

In the course of our study several special `anti-resonant' cases
 arise. One of these occurs in the $N=4$ clock hamiltonian, as defined in Ref.
\onlinecite{Clarke2013} and closely related to the Ashkin-Teller model (see e.g.
Ref.~\onlinecite{Kohmoto1981}). We show that this model can be rewritten as a
frustrated spin-$1/2$ ladder in such a way that the couplings along the rungs
vanish precisely at the achiral resonant points. Here the system is exactly
solvable and can be mapped to a $\ZZ_4$ time-reversal invariant Majorana model,
i.e. two uncoupled Majorana chains. At these points, the model also displays
exact $\ZZ_4$ parafermionic zero-modes which can be expressed in terms of the
Majorana modes of the two decoupled chains.

Another interesting special case is the region around the $\theta= \pi/6$ chiral
point of the $N=3$ model, where an apparent resonance is effectively cancelled by
a precise match up of contributing domain-wall excitation energies. This region
was pinpointed in earlier studies of the problem\cite{Jermyn2014} as being a promising candidate
for a strong zero-mode. Although our results show the special region around the
$\pi/6$ point in parameter space again becomes vanishingly small in the limit $L
\rightarrow \infty$, our analytic and numerical results suggest that
strong zero-modes do exist at the point itself. We note furthermore that when
the universal degeneracy breaks down in the nearby parameter space, it typically
does so at very high energies.

An outline of the paper is as follows. In section \ref{sec:the_model}, we
present the model and rewrite it using a domain-wall picture which is useful for
the subsequent analysis. In section \ref{sec:unperturbed_system} we discuss the
unperturbed system and characterise the resonant crossings between bands. In
section \ref{sec:perturbation_theory} we analyse the effect of perturbations at
the resonant crossings and also far away from any crossings. In section
\ref{sec:ED} we discuss the results of our exact diagonalisation analysis and
compare them to the aforementioned perturbative analysis. In section
\ref{sec:antires-hidden} the recovery of zero-modes at anti-resonant points, the
exactly solvable nature of the $N=4$ model at achiral points and hidden
zero-modes are discussed. Finally we wrap up with the conclusions and outlook in
section \ref{sec:conclusions}.

\section{The Model}\label{sec:the_model}

\subsection{Clock Spins and Parafermions}

We consider the $\ZZ_{N}$ quantum clock hamiltonian as given in Refs:
\onlinecite{Clarke2013,Jermyn2014} and written $H=H_J+H_f$, where
\begin{eqnarray}
\label{eqn:hamiltonian}
H_J &=& - J e^{i \theta} \sum_{x=1}^{L-1} \sigma_x^{\dagger} \sigma_{x+1}  + h.c. \\  \non
H_f &=&-  f e^{i \phi} \sum_{x=1}^{L} \tau_x    + h.c.
\end{eqnarray}
$N$ is the number of 'hour-marks' on the clock at each site. The operators $\sigma$
and $\tau$ are $N\times N$ matrices defined by 
\[
\sigma_{i,j} = \omega^{i-1}\delta_{i,j} ~~~~~~
\tau_{i,j} = \delta_{i+1,j \Mod{N}},
\]
with $\omega = e^{\frac{2i \pi}{N}}$. The operator $\sigma$ measures
the clock value at a given site. The operator $\tau$ reduces the clock value on a given
site by one. For $N=2$, $\sigma$ and $\tau$ are the $\sigma^z$ and $\sigma^x$ Pauli
operators respectively. For $N=3$, they take the form
\begin{equation}
\sigma=\left[ \begin{array}{ccc}
1 & 0 & 0  \\
0 & \omega & 0  \\
0 & 0  & \omega^2
\end{array} \right]
,\quad
\tau=\left[ \begin{array}{ccc}
0 & 1 & 0  \\
0 & 0 & 1  \\
1 & 0 & 0
\end{array} \right]
\label{eqn:sigma_tau_N_3}
\end{equation}
Note that for $N=2$, this model is the transverse field Ising spin chain. (In
this case the angles $\theta$ and $\phi$ are redundant and can be removed by
dividing $J$ and $f$ by real factors $\cos(\theta)$ and $\cos(\phi)$). When
$N>2$ and $\theta$ is not a multiple of $\frac{2\pi}{N}$, the model is called
the chiral clock model since spatial-parity and time reversal symmetries are
broken\cite{Fendley2012}. For $N=3$ this model is equivalent to the clock models
analysed in Ref.~\onlinecite{Fendley2012}, while for higher $N$, it is a special
case of the general $\ZZ_N$ models of in Ref.~\onlinecite{Fendley2012}. We
consider the regime with $f \ll J$ which is the ordered phase. This can be
mapped exactly to the topologically ordered phase of the corresponding
parafermion chain model. The $\phi$ angle does not have much of an effect on
many features and we will often just consider the $\phi=0$ case. There is one
special feature at $\phi=0$, which is the presence of dihedral symmetry -- see
Sec.~\ref{sec:dihedral}

The operators $\tau$ and $\sigma$ are $N^{\text{th}}$ roots of the identity
\begin{equation}
\sigma^N = \tau^N =I
\end{equation}
and satisfy
\begin{equation}
\tau_x \sigma_y = \omega^{\delta_{x,y}} \sigma_y \tau_x
\end{equation}
Using this, one may see that the hamiltonian possesses a global symmetry: the
operator $Q$ which moves the clock back one click at all sites,
\begin{equation}
Q=\prod_{i=1}^{L} \tau_x.
\label{eqn:Qop}
\end{equation}
The eigenvalues of Q are $\omega^q$ with $q=0,1,...N-1$. We call the different
eigenspaces of the $Q$ operator the $q$-sectors and label each with its respective $q$
value.

Using a non-local transformation due to Fradkin and Kadanoff\cite{Fradkin80} and
analogous to the Jordan--Wigner transformation, we can rewrite the spin model in
terms of parafermionic variables. Parafermionic operators are defined as
follows,
\[
\gamma_{2x-1}=\sigma_{x}\prod_{j<x}\tau_{j} ~~~~~~ \gamma_{2x}=\omega^{(N-1)/2}\sigma_{x}\prod_{j\le x}\tau_{j}
\]
These satisfy the relations
\begin{equation}
\gamma_{x}\gamma_{y}=\omega^{\mathrm{sgn}(y-x)}\gamma_{y}\gamma_{x}
~~~~~~~~ \gamma_{x}^{N}=1
\label{eq:paraf_alg}
\end{equation}
The factor of $\omega^{(N-1)/2}$ in $\gamma_{2x}$ is necessary for the second
equation. One now finds that
\begin{eqnarray}
H_{J}&=&-J\omega^{(N-1)/2}e^{i\theta}\sum_{x=1}^{L-1}\gamma^{\dagger}_{2x}\gamma^{\phantom{\dagger}}_{2x+1} + h.c. \non \\
H_{f}&=&-f\omega^{-(N-1)/2}e^{i\phi}\sum_{x=1}^{L}\gamma^{\dagger}_{2x-1}\gamma^{\phantom{\dagger}}_{2x} + h.c.
\end{eqnarray}
and 
\[Q=\omega^{-L(N-1)/2}\prod_{x=1}^{L}\gamma^{\dagger}_{2x-1}\gamma^{\phantom{\dagger}}_{2x}
\]
In these variables, $Q$ is a topological quantum number analogous to the
fermionic parity $(-1)^{F}$. In fact $\gamma_x Q=\omega Q \gamma_x$ and we see
that $Q$ will commute only with operators constructed from products of $N$
parafermion operators. $Q$ is called $N$-ality, or sometimes ``parafermionic
parity''. When $f=0$, the hamiltonian does not contain the operators $\gamma_1$
and $\gamma_{N}$ and these operators are then strong edge zero-modes, giving
rise to an $N$-fold degeneracy in the entire spectrum. Of course in the spin
model we can understand this degeneracy as a trivial case of spontaneous
breaking of the $\ZZ_N$ spin symmetry since there are no spin flip terms in the
hamiltonian at $f=0$. Note that at $f=0$, the operators $\sigma_x$ will in fact
act as zero-modes for all $x$. However, for $x$ away from the edges, these
operators are non-local in the parafermion language. Also, as $f$ is taken to
nonzero values, only the edge zero-modes have a chance to persist as strong
zero-modes. In the rest of the paper, we will focus on determining to what
extent the degeneracies of the model persist when $f\neq 0$.

\subsection{$Q$-eigenstates}
The hamiltonian (\ref{eqn:hamiltonian}) is written in the position space clock basis, where
states are written $\ket{\textbf{s}}^S = \ket{s_1,s_2,... ,s_L}^S$, and
$s_x\in\{0,1,..., N-1\}$ is the clock value on site $x$, or in other words
$\sigma_x \ket{s_1,s_2,... ,s_L}^S= \omega^{s_x}\ket{s_1,s_2,... ,s_L}^S$. The
superscript $S$ explicitly indicates position space.



These states are not eigenstates of the $Q$ operator (\ref{eqn:Qop}), but are
super-positions of eigenstates from different $q$-sectors. The action of $Q$ on
$\ket{\textbf{s}}^S$ changes the value of the clock at each site by $-1
\Mod{N}$.
\begin{equation}
Q\ket{\textbf{s}}^S = \ket{\textbf{s} - 1}^S
\end{equation}
For each state $\ket{\textbf{s}}^S$, we can write down a state in a given
$q$-sector by performing a discrete Fourier transform as in
\begin{equation}
\label{eq:QNFourier}
\ket{\textbf{s}}^{S}_{q}\sim\frac{1}{\sqrt{N}} \sum_{j=0}^{N-1} \omega^{-jq} Q^j\ket{\textbf{s}}^S,
\end{equation}
which is an eigenstate of $Q$, with eigenvalue $\omega^{q}$. For what follows it
is necessary to choose a convention for the global phase of these states.
Here it will be convenient to fix the phase according to the value of the clock
on the last site. The component of the state where the last site has its clock
set to zero will be taken to have a positive real coefficient. Since $\sigma_L$
measures the clock value at site $L$, this can be achieved using the
definition
\begin{equation}
\label{eq:QNFourier_sigma}
\ket{\textbf{s}}^{S}_{q} \equiv \frac{1}{\sqrt{N}} \sum_{j=0}^{N-1} \sigma_L^{q} Q^j\ket{\textbf{s}}^S.
\end{equation}

\subsection{Domain-wall picture}
\label{sec:domain_wall_picture}
We now introduce a domain-wall representation where one focuses on the
differences between clock values on neighbouring sites, rather than the clock
values themselves. This picture allows us to directly write down the hamiltonian
in each $q$-sector. We find that by doing this, and as a consequence of the
phase convention we chose in the previous section, there are only two terms in
the hamiltonian which depend on the particular $q$-sector. Better still, both of
these act at the right end of the chain. This makes the domain-wall
representation very convenient for numerical calculations and a natural setting
for perturbative expansions.

Each $q$-sector is spanned by $(\mathbb{C}^N)^{\otimes(L-1)}$ states which we label
$\ket{\textbf{d}}^D_q = \ket{d_1, d_2,...,d_{L-1}}$ where $d_x\in\{0,1,...,N-1\}$.
$\ket{\textbf{d}}^D_q$ is a sum over position spin space states like in
(\ref{eq:QNFourier_sigma}), where $\textbf{s}$ satisfies $d_x=s_{x+1}-s_{x}
\Mod{N}$, for $1 \le x \le L-1$. We illustrate this with an explicit example for
a state with $N=3$ and $L=4$ for all $q$.
\begin{eqnarray}
\ket{012}_0^D &=& \ket{0010}^S + \ket{1121}^S + \ket{2202}^S \nonumber \\
\ket{012}_1^D &=& \ket{0010}^S + \omega \ket{1121}^S + \omega^2 \ket{2202}^S \nonumber\\
\ket{012}_2^D &=& \ket{0010}^S + \omega^2 \ket{1121}^S + \omega \ket{2202}^S
\end{eqnarray}
We now proceed to write the hamiltonian (\ref{eqn:hamiltonian}) in this
representation. First we define operators $\alpha$ and $\beta$ given by
\begin{eqnarray}
\alpha_{x} &=& \prod_{i=1}^{x}\tau_x \nonumber\\
\beta_{x} &=&  \sigma_{x}^{\dagger} \sigma^{\phantom{\dagger}}_{x+1}
\end{eqnarray}
These operators have the same matrix representations as operators $\tau$ and
$\sigma$ respectively, except that they act on the domain-wall basis. In terms
of these operators $\tau_{x}$ can be written as
\begin{eqnarray}
  \tau_x =
    \begin{cases}
      \alpha_1 & x = 1 \\
      \alpha_{x-1}^{\dagger}\alpha_x & 1 < x < L \\
      Q\alpha_{L-1}^{\dagger} & x = L
    \end{cases}
\end{eqnarray}
and the hamiltonian (\ref{eqn:hamiltonian}) is now given by
\begin{eqnarray}
\label{eq:Halpha}
H_{J} &=& -J e^{i\theta}\sum_{x=1}^{L-1} \beta_x +h.c \\
 H_{f}  &=& - f e^{i\phi}  \left( \alpha_1   + \sum_{x=1}^{L-2} \alpha_x^{\dagger}\alpha_{x+1} +Q \alpha_{L-1}^\dagger  \right)+ h.c.
\end{eqnarray}
Note that in this form all $q$-dependence appears only in $H_{f}$ on terms
$-fe^{i\phi}Q \alpha_{L-1}^\dagger$ and $-fe^{-i\phi}\alpha_{L-1}Q^\dagger$ 
which act on the end of the chain.

Since $Q$ commutes with $H$ and with
$\alpha^{\phantom{\dagger}}_{L-1}$ and $\alpha^{\dagger}_{L-1}$, this
hamiltonian is block diagonal, and we can write the hamiltonian in each
$q$-sector by replacing $Q$ with the appropriate eigenvalue,
\begin{eqnarray}
\label{eq:Halpha_q}
H_{J}^q &=& -J e^{i\theta}\sum_{x=1}^{L-1} \beta_x +h.c \\
 H_{f}^q  &=& - f e^{i\phi} \left( \alpha_1   + \sum_{x=1}^{L-2} \alpha_x^{\dagger}\alpha_{x+1} + \omega^q\alpha_{L-1}^\dagger  \right)+ h.c.
\end{eqnarray}

For later sections it will be useful to decompose $H_f$ further into terms that
act on the edge and those containing bulk terms. This gives $H_f^q = H_{f_e}^q +
H_{f_b}^{q}$ where
\begin{eqnarray}
 H_{f_e}^{q}  &=& - f e^{i\phi} \left( \alpha_1  + \omega^q \alpha_{L-1}^\dagger\right)  + h.c.\\
 H_{f_b}^{q}  &=& - f e^{i\phi} \sum_{x=1}^{L-2} \alpha_{x}^\dagger \alpha_{x+1}  + h.c.
\label{eqn:Hf_edge}
\end{eqnarray}

We can now interpret $(\alpha^{\dagger}_x)^k$ as creating a domain-wall of type $k$ at site
$x$. For arbitrary $\ket{\textbf{d}}_q^D$ we may then write
\begin{equation}
\ket{\textbf{d}}_q^D = \prod_{x=1}^{L-1} (\alpha_x^{\dagger})^{d_x}\ket{\emptyset}_q
\end{equation}
where $\ket{\emptyset}_q = \ket{0,0...0}_q^D$ is the state without domain-walls
and with the appropriate eigenvalue of $Q$. As an example we write
\begin{equation}
\ket{012}_q^D = \alpha_2^{\dagger} \alpha_3^{\dagger 2} \ket{\emptyset}_q
\end{equation}
While this way of writing the states is reminiscent of second quantization, one
should remember that the "vacuum states" $\ket{\emptyset}_q^D$ are not annihilated by
the domain-wall annihilation operators and also that
$[\alpha^{\phantom{\dagger}}_{x},\alpha^{\dagger}_{y}]=0$, so the domain-wall
creation operators are not the usual creation operators of bosonic or fermionic
Fock space. There have been attempts to formulate a Fock space for
parafermions, notably the work of Cobanera and Ortiz \cite{Cobanera2014}.

\subsection{Dihedral symmetries}
\label{sec:dihedral}

When the angle $\phi$ is a multiple of $\frac{2\pi}{N}$, the models we study
here have an enlarged group of symmetries, isomorphic to $D_N$, the symmetry
group of a regular polygon with $N$ sides. The $\ZZ_N$ generated by $Q$ is
included in this $D_N$ as the group of rotations of the polygon. The extension
of the symmetry group from $\ZZ_{N}$ to $D_{N}$ is not generally present in the
models of Ref.~\onlinecite{Fendley2012} for $N>3$ (even at $\phi=0$). Since
$D_{N}$ is a non-Abelian group, its presence causes degeneracies in the
spectrum. In particular, as long as $\phi=0$, any eigenstate of $Q$ with
eigenvalue $\omega^{q}$ such that $\omega^{q}\neq\omega^{-q}$ comes with a
partner at the same energy whose $Q$ eigenvalue is $\omega^{-q}$.

We now present the $D_{N}$ symmetry explicitly\footnote{Our presentation of the
$D_{N}$ symmetry group here is in accordance with the results given by Motruk,
Berg and Pollmann\cite{Motruk2013} and generalizes some of their findings.}.
First of all recall that the dihedral group $D_{N}$ is generated by two elements
$a$ and $b$ subject to the relations $a^{N}=b^2=(ab)^2=e$ (where $e$ is the
group identity). Here, $a$ can be viewed as a rotation over an angle $\frac{2
\pi}{N}$ and $b$ as a reflection acting on the regular $N$-gon. The group has a
total of $2N$ elements, the rotations $a^{j}$ and the reflections $a^{j}b$, with
$j\in\{0,...,N-1\}$. We now introduce an operator $R$, such that
$R^2=(QR)^2=I$, which together with $Q$ (from (\ref{eqn:Qop})) generates a
$D_{N}$ symmetry. Next we define $\eta$ to be the operator that transforms each
clock value as
\[
\eta\ket{s}=\ket{N-s}
\]
with $\eta\ket{0}=\ket{0}$. Note that $\eta$ corresponds to the complex
conjugation of the eigenvalues of $\sigma$. We see that $\eta^{2}=1$ and that
the following exchange relations with $\tau$ and $\sigma$ hold
\begin{equation}
\label{eq::sigma&theta}
\sigma\eta=\eta\sigma^\dagger ~~~~~ \tau\eta=\eta\tau^\dagger
\end{equation} 

We can now define $R_0=\prod_{i}\eta_i$. This satisfies $(R_0)^2=I$, $(R_0
Q)^2=1$ and $R_0 H(\theta,\phi)=H(-\theta,-\phi) R_0$. We note that $R_0$ and
$Q$ generate a $D_N$ group together, but $R_0$ is only a symmetry if
$\theta=\phi=0$. Combining $R_0$ with the operator $F$ which flips the clock
state, that is, which switches the clock state at position $i$ on the chain with
the clock state in position $L-i$, we obtain an operator $R=R_0 F$ which is a
symmetry for all $\theta$ as long as $\phi=0$. Explicitly, we define
$F\ket{s_1,s_2,...,s_{L}}^{S}=\ket{s_{L},...,s_{2},s_{1}}^{S}$ and $R=R_0 F$. We
then find that $R^2=(RQ)^2=I$ and
\[
R H( \theta, \phi)=H(\theta,-\phi)R,
\]
which in particular gives $[H(\theta,\phi=0),R]=0$. We see that we have
$D_{N}$-symmetry at $\phi=0$ and in fact, $\ZZ_2\times D_{N}$ symmetry if
$\phi=\theta=0$, as we have $F$ as an additional commuting symmetry when
$\theta=0$.

To see how this leads to degeneracies, note that $QR=RQ^{-1}$ which means that
if $\ket{\psi}$ is an eigenstate of $H$ and of $Q$ with eigenvalues
$(E,\omega^{q})$ then $R\ket{\psi}$ is an eigenstate of $H$ and $Q$ with
eigenvalues $(E, \omega^{-q})$. As a result all eigenstates of $Q$ with
eigenvalues $\omega^{q}\neq\omega^{-q}$ are at least doubly degenerate. Only
states with real eigenvalues of $Q$ have a chance of being nondegenerate.
Generically only states within a given irreducible representation of $D_{N}$
will be degenerate (barring the existence of some further symmetry but we don't
observe this). Since $D_{N}$ has only one dimensional and two dimensional
irreducible representations for all $N$ (see for instance
Ref.~\onlinecite{james_and_liebeck}) we see that the $D_{N}$ symmetry only
causes doublets and not higher multiplets of degenerate states (or bands) to
appear. We find that the $D_{N}$ doublets are split when $\phi\neq 0$, so it is
not possible to extend the $D_{N}$ symmetry to this regime.

\section{Unperturbed system}\label{sec:unperturbed_system}
The regime of interest for topologically non-trivial phases in these systems is
where $f \ll J$. In this work we will explore this regime by performing
perturbative expansions in $H_f^q$. Before doing this however, we will spend
some time exploring the unperturbed system, when $f=0$ which will serve to guide
our later analysis.

\subsection{Unperturbed energy levels}

The unperturbed system $H_J$, is diagonal in both representations 
discussed in the previous section. In the domain-wall picture, the energy of a
state $\ket{\textbf{d}}_q^D$ has the simple form
\begin{equation} E = -2J\sum_{x=1}^{L-1} \cos(\theta + \frac{2\pi d_x}{N}) =
\sum_{x=1}^{L-1}\epsilon_{d_x}
  \label{eqn:DWH0Energy}
\end{equation}
where we recall that $d_x =s_{x+1}-s_x$. We also define $\epsilon_j \equiv
-2J\cos(\theta + \frac{2\pi j}{N})$ for compactness. There is no $q$ dependence
in the unperturbed hamiltonian which means there is an exact $N$ fold degeneracy
for the entire spectrum, consistent with the presence of the zero-modes at $f=0$.

\begin{figure}
\includegraphics[width=0.45\textwidth]{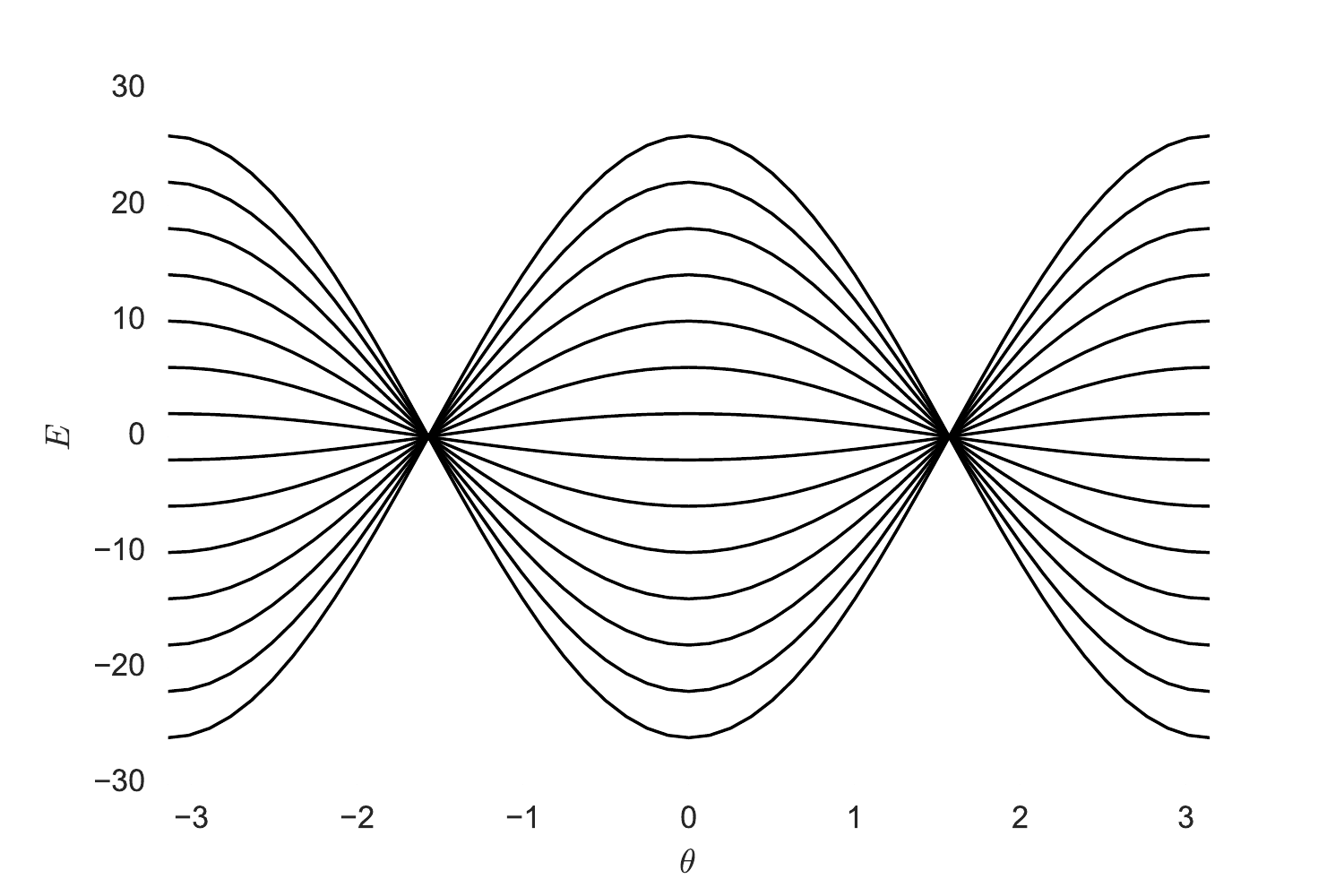}
\includegraphics[width=0.45\textwidth]{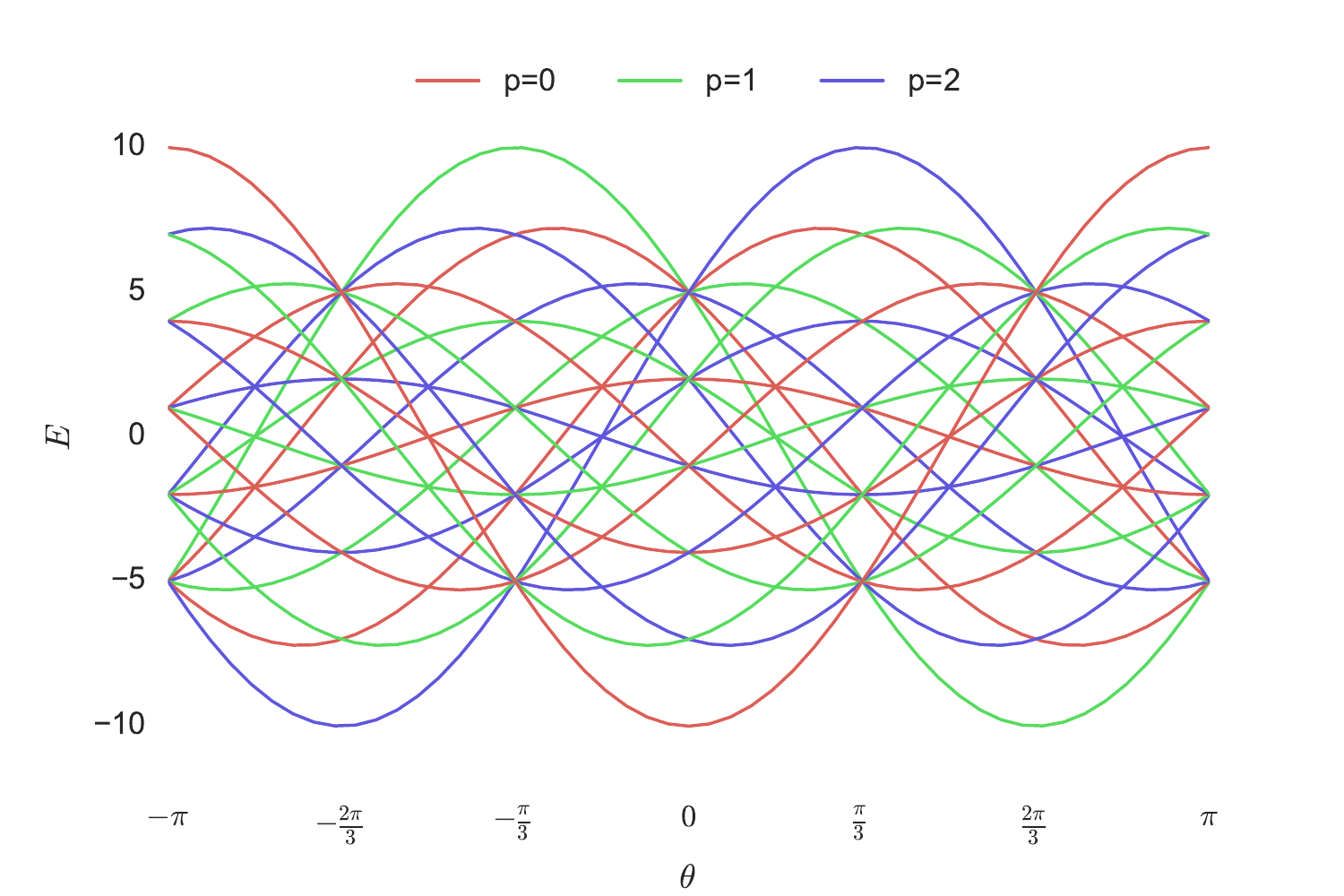}
\includegraphics[width=0.45\textwidth]{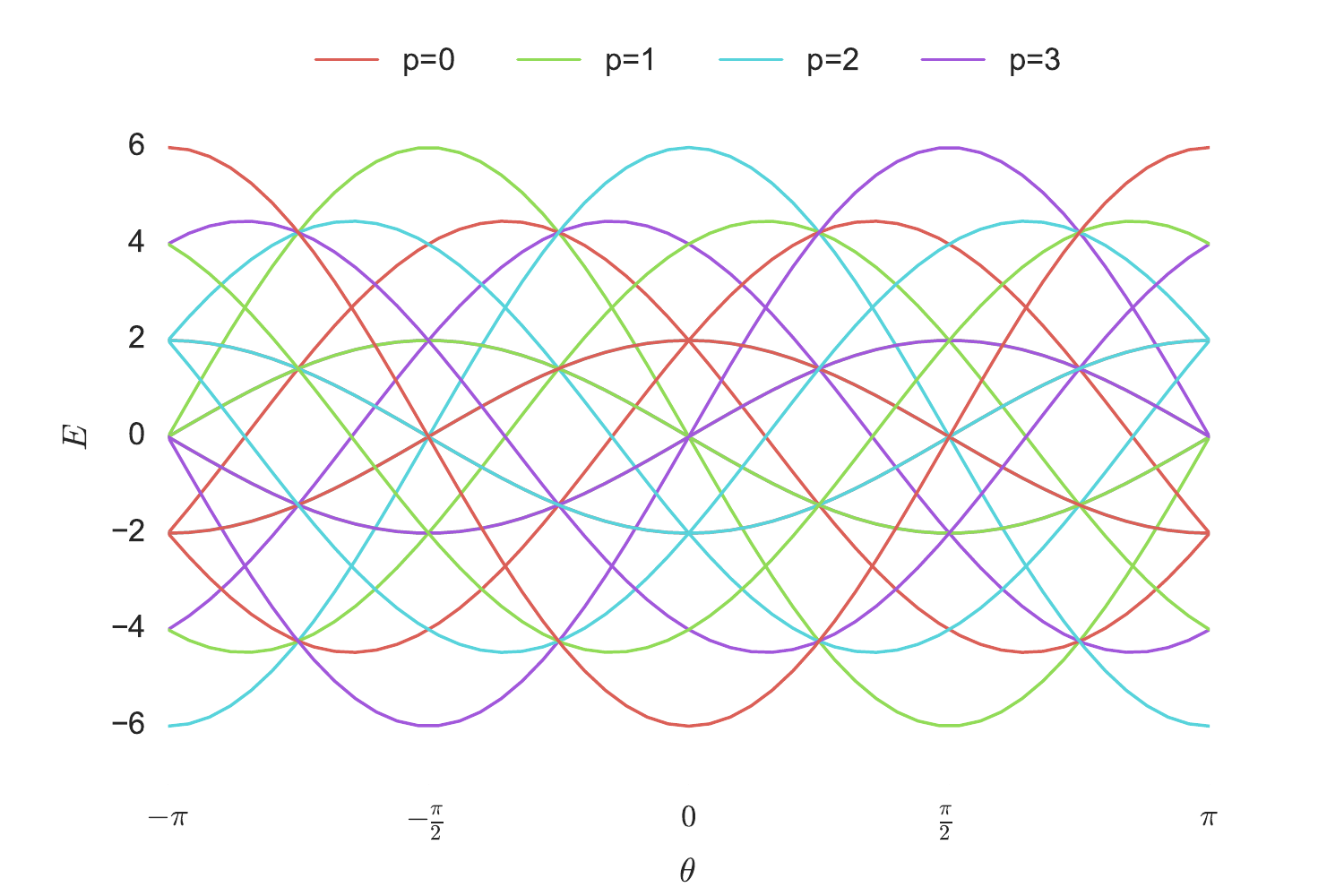}
\caption{\emph{Top:} Spectrum of unperturbed  $N=2$ chain with $L=14$.\\
\emph{Middle:} Spectrum of unperturbed  $N=3$ chain with $L=6$.\\
\emph{Lower:} Spectrum of unperturbed  $N=4$ chain with $L=4$.}
\label{fig:N234Spectrum}
\end{figure}

Figure \ref{fig:N234Spectrum}
shows the energy spectra of the unperturbed system plotted against $\theta$ for
$N=2,3$ and $4$. Despite the fact that these spectra are for very small chains,
they capture the important features.

There are degenerate bands of states, with each band labeled by the number of
each type of domain wall present. In the absence of $H_f$, the domain walls do
not disperse and hence there is a large degeneracy associated with different
choice of the domain wall positions. A further useful label for the bands, which
is also indicated with colour in the figures, is defined by
\[
p=s_L - s_1 =\sum_{x=1}^{L-1}d_x~~~ (\mathrm{mod}~N)
\]
We can think of $p$ as the total number of domain-walls modulo $N$. This
quantity, which we call total domain-wall angle, is crucial to understanding the
perturbative analysis discussed in later sections. The underlying symmetry $P$
of $H_J$ can be written as
\begin{equation}
  P=\sigma_1^{\dagger} \sigma_L= \prod_{x=1}^{L-1}\beta_x.
  \label{eqn:DWParityOp}
\end{equation}
and has eigenvalues $\omega^p$. 

At $\theta=0$, the $N$ fold degenerate ground state is 'ferromagnetic' and
spanned by states with all neighbouring clocks aligned. The ground state in each
q-sector is written $\ket{00...0}^D_q$. At $\theta=\frac{2\pi k}{N}$, for
nonzero integer $k$, the spectrum is identical, but ground states are
 spanned by states with all neighbouring clocks $k$
clock positions apart,
written $\ket{kk...k}^D_q$. It is interesting to note that for $N$ odd, the
spectrum is not symmetric around $E=0$, except at $\theta = \frac{k\pi}{2N}$,
with $k$ odd, where the system also has its ``superintegrable'' point for $N=3$ \cite{Fendley2012}.

As noted, the energy of a band in the unperturbed system depends only on the
number and type of domain-walls, but not their location. If we count the trivial
domain walls where adjacent clock variables are equal, then every pair of
neighbouring sites gives rise to a domain wall and we have $L-1$ domain walls of
$N$ distinct types. The maximum number of unique energy levels for a chain of length $L$
is then the number of ways of distributing the $L-1$ domain-walls among $N$
numbered boxes. This is the number of compositions of $L-1$ with $N$ parts,
denoted $c_S(L-1, N)$ \cite{Eger2013}. $S$ here is the set of possible values
the parts can take, so here $S=\{0,...,L-1\}$. The $c_S(L-1, N)$ are extended
binomial coefficients \cite{Eger2013} and satisfy the recurrence relation
\begin{equation}
c_S(L-1, N) = c_S(L-2, N) + c_S(L-1, N-1)
\end{equation}
which can be used to calculate these efficiently for large systems. 

Each composition is an ordered tuple of non-negative integers $(p_0, p_1,...,p_{N-1})$
that satisfy 
\begin{equation}
\sum_{j=0}^{N-1} p_j = L-1.
\label{eqn:diophantine_eqn}
\end{equation}
We use these tuples to label the bands of states of the unperturbed system. Here
$p_j$ is the number of domain-walls of type $j$ in the band. The energy of a
band in the unperturbed system described by $\vec{p}$ can be written
\begin{equation}
E_{\textbf{p}}(\theta) =  -2J\sum_{j=0}^{N-1} p_j\cos(\theta + \frac{2\pi j}{N}) = \sum_{j=0}^{N-1}p_j\epsilon_{j}
\label{eqn:unperturbed_energy}
\end{equation}

\subsection{Mapping of resonance points}\label{sec:mapping_resonances}
We define a resonance point to be a place where two bands of the unperturbed
system cross and become degenerate. At and in the vicinity of resonance points
all crossing bands must be considered when performing perturbative expansions.
It is then of interest to ask for which $\theta$ values these resonance points
occur, and whether they are dense on the $\theta$-axis in the limit
$L\rightarrow \infty$. In this section we derive expressions for the $\theta$
values at which resonance points occur and following this show that for $N>2$
they are dense on the $\theta$-axis.

Let us consider two bands, labelled by tuples $\vec{a}$ and $\vec{b}$ where $a_j$
gives the number of domain-walls of type $j$ for the first band, etc. We write
$\vec{c}=\vec{a}-\vec{b}$, and note that $\vec{c}$ satisfies the following
constraints:
\begin{eqnarray}
\label{eq:difconstr}
\sum_j^{N-1}c_j&=&0 \nonumber\\
0\le n_{\vec{c}}&:=&\sum_j^{N-1}|c_j|\le 2(L-1) ~{\rm with~} n_{\vec{c}} \in 2\ZZ
\end{eqnarray}
The quantity $n_{\vec{c}}$ is useful as it is related to the order in
perturbation theory necessary to connect states in different bands at a
resonance point.

\begin{figure}
\includegraphics[width=0.45\textwidth]{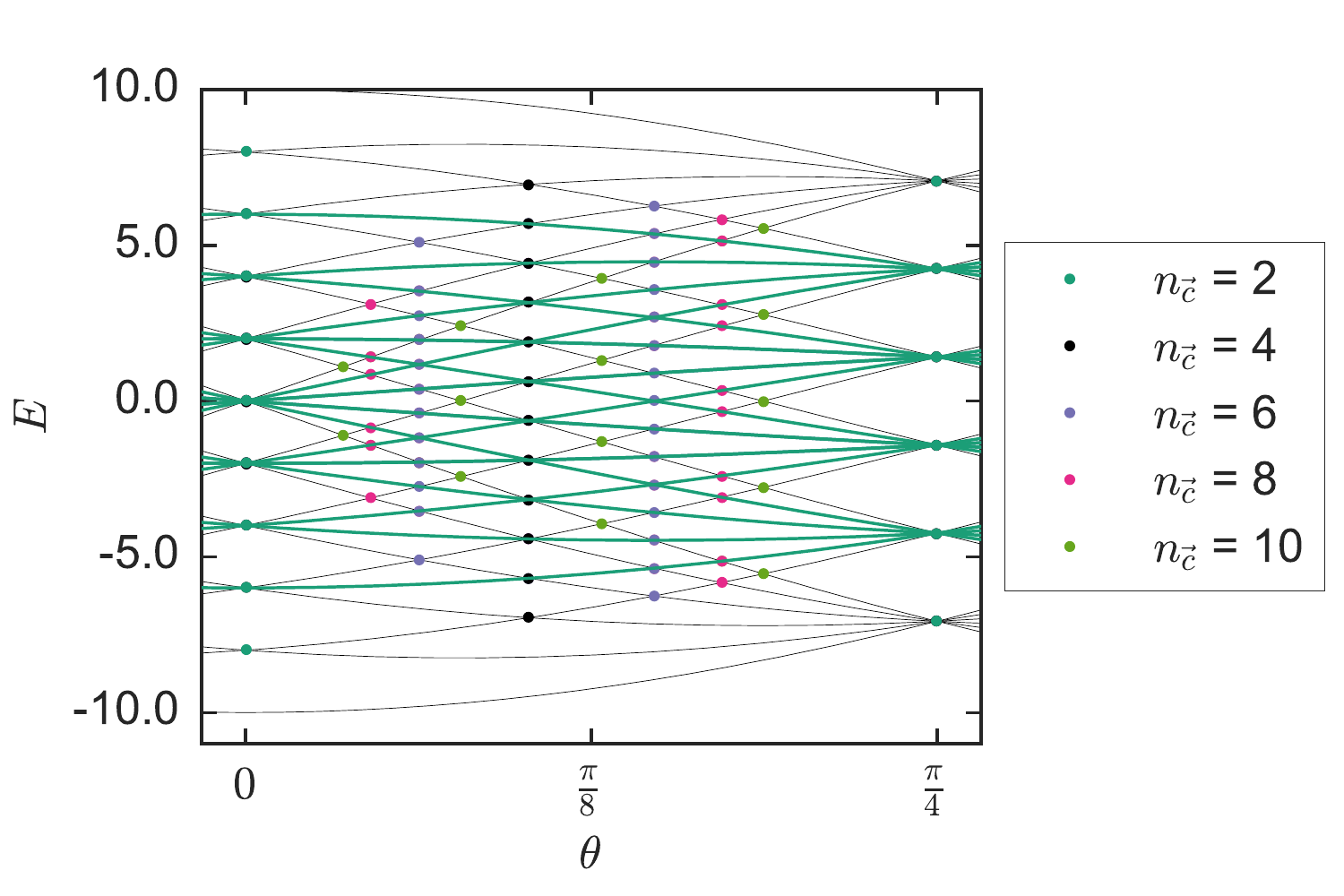}
\caption{Unperturbed spectrum of $N=4$ system of length $L=6$ with resonance
  point highlighted. Each resonance points is labelled by a number
  $n_{\vec{c}}$, which is related to the order on which perturbing terms can
  couple the bands at that point. Where multiple resonance points occur at the
  same location, the one with the lowest $n_{\vec{c}}$ is highlighted. Some
  bands in this case are degenerate everywhere which are indicated by the
  green lines.}
\label{fig:N4L8Map}
\end{figure}

\begin{figure}
\includegraphics[width=0.45\textwidth]{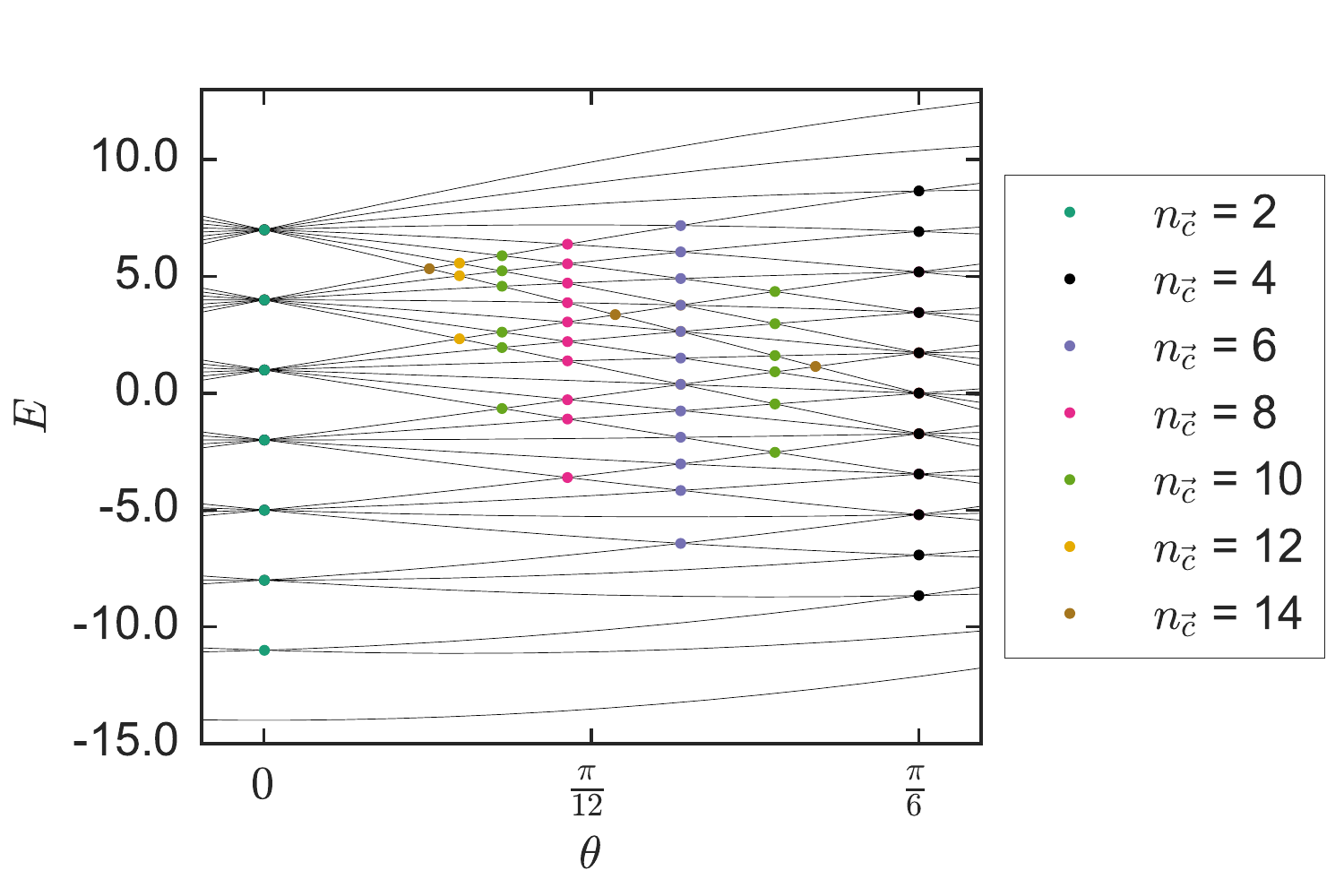}
\caption{Unperturbed spectrum of $N=3$ system of length $L=8$ with resonance
  point highlighted. Each resonance points is labelled by a number
  $n_{\vec{c}}$, which is related to the order on which perturbing terms can
  couple the bands at that point. Where multiple resonance points occur at the
  same location, the one with the lowest $n_{\vec{c}}$ is highlighted.}
\label{fig:N3L8Map}
\end{figure}

From (\ref{eqn:unperturbed_energy}), we see that the bands labelled by $\vec{a}$
and $\vec{b}$, are degenerate precisely when
\begin{equation}
\label{eq:banddegen}
\cos(\theta) \sum_j^{N-1}c_j\cos(\frac{2\pi j}{N})= \sin(\theta)\sum_j^{N-1}c_j\sin(\frac{2\pi j}{N})
\end{equation}
Assuming that both sides of this equation are nonzero, we find that 
the bands are degenerate at $\theta$ values which satisfy
\begin{equation}
\tan(\theta) = \frac{\sum_j^{N-1}c_j\cos(\frac{2\pi j}{N})}{\sum_j^{N-1}c_j\sin(\frac{2\pi j}{N})}
\label{eqn:phi_at_crossing}
\end{equation}
and hence the bands cross twice over the full range of possible $\theta$ values.
Alternatively, it may be that both sides of (\ref{eq:banddegen}) are zero. In
this case the bands are degenerate for all $\theta$ and we can write (still with
$\omega=e^{2\pi i/N}$)
\begin{equation}
\sum_{j}c_{j}\omega^{j}=0.
\label{eqn:everywhere_degenerate_constraint}
\end{equation}
When $N=3$, it is easy to see that this can never be satisfied with
$\sum_{j}c_{j}=0$ and $\vec{a} \ne \vec{b}$. Hence there are no bands which
are everywhere degenerate for $N=3$. However for $N=4$, there are nontrivial
difference vectors $\vec{c}$ which satisfy equation
(\ref{eqn:everywhere_degenerate_constraint}) and the constraints given in
Eq.~(\ref{eq:difconstr}); the vectors $\vec{c}=(p,-p,p,-p)$
give solutions for all integer $p$ such that $4|p|\le 2(L-1)$. This leads to the
appearance at $N=4$ of precise degeneracies between bands, at all $\theta$, that
comprise of a different number and type of domain-walls. Figure
\ref{fig:N4L8Map} shows the distribution of the bands (in green) that are
everywhere degenerate (e.g. bands with different domain-wall composition lie
on top of each other everywhere) and the resonance points and for a $L=6$ system
with $N=4$. A contrasting plot for $N=3$, Figure \ref{fig:N3L8Map}, shows that
bands with different number and type of domain-walls typically have different
energies. The exception in this case of course being at the resonant crossings.
These examples reflect the general situation for any $N$, with
everywhere-degenerate bands appearing always when $N$ is composite and never
when $N$ is prime.


Fully characterising the degenerate bands comes down to a characterisation of
all linear relations of $N^{\rm th}$ roots of unity with integer coefficients. This is an interesting
problem in number theory -- some recent results and references can be found in
Ref.~\onlinecite{Steinberger2008}. The essential difference between prime and
composite $N$ can be easily seen however. First of all, if $N$ is prime, then
all nontrivial linear combinations of the $N^{\rm th}$ roots of unity with
integer coefficients satisfying equation
(\ref{eqn:everywhere_degenerate_constraint}) have constant coefficients. This
means $c_j=p$ for some $p\in\ZZ$ for all $j$ and we see that the constraint
$\sum_{j}c_{j}=0$ is never satisfied. Hence when $N$ is prime there are never
any bands that are degenerate for all $\theta$. We now show why this is the case
for $N$ prime, but not for $N$ composite. For $N$ prime the cyclotomic
polynomial $1+x+...+x^N$ is the minimal polynomial for the primitive roots
$\omega^j$ (for $j\neq 0$). If there were a linear combination of the powers of
$\omega$ with non-constant coefficients, such that
$\sum_{j=0}^{N-1}d_{j}\omega^{j}=0$ then we could obtain a nontrivial linear
combination which omits $\omega^{N-1}$, namely
$\sum_{j=0}^{N-2}(d_{j}-d_{N-1})\omega^j=0$. However, this would show that
$\omega$ is the root of the polynomial $\sum_{j=0}^{N-2}(d_{j}-d_{N-1})x^j$
which is of lower degree than the minimal polynomial leading to a contradiction.
For composite $N$, we can always choose integer factors $e$ and $f$ of $N$, and
find non-trivial coefficients $\{d_j\}$ which satisfy
$\sum_{j=0}^{e-1}d_{j}\sum_{k=0}^{f-1}\omega^{ek+j} = 0$ and
$\sum_{j=0}^{e-1}d_{j}=0$. These satisfy the
constraint that the sum of coefficients vanish and that these combinations all
equal zero since $\sum_{k=0}^{f-1}\omega^{ek + j}=0$ forall $0 \le j \le e$.
Hence we have everywhere-degenerate bands for all composite $N$.

\subsection{Resonance points are dense on the $\theta$ axis as $L \rightarrow \infty$}

We can now deal with the question of whether resonance points occur
arbitrarily near any particular value of $\theta$. This is clearly not the case
for finite $L$, but as $L$ grows, ever more resonance points appear and in the
limit $L\rightarrow\infty$ they do form a dense set on the $\theta$-axis, for
all $N>2$. To see this we first note that the two solutions to
Eq.~(\ref{eqn:phi_at_crossing}) are in fact given by
\[
\theta =\frac{\pi}{2}- \arg\left(\sum_{j} c_j \omega^{j}\right) ~~({\rm mod~} \pi).
\] 
It is clear that we can make the linear combination $\sum_{j} c_j \omega^{j}$
take any fixed argument $\alpha$ to arbitrary accuracy if we are allowed to take
the coefficients $c_j$ as large as we want. However a complication is once again
the condition that $\sum_{j}c_j=0$. We can circumvent this as follows. First of
all note that we can choose any two powers of $\omega$, for example
$\omega$ and $\omega^0=1$ as a basis for the complex plane over the reals. We
can then take a point $z$ with $\arg(z)=\alpha$ in the complex plane and write
$z=a+b\omega$ with $a,b\in \RR$. The coefficients $a,b\in \RR$ can be
arbitrarily well approximated by rationals $a',b'\in \QQ$, in particular, for
any chosen $\epsilon$, we can make sure that
$|\arg(a'+b'\omega)-\alpha|<\epsilon$. We can now write $a'=\frac{a_1}{a_2}$ and
$b'=\frac{b_1}{b_2}$ for integers $a_1,a_2,b_1,b_2$ and consider $Z=N a_2 b_2
(\frac{a_1}{a_2} + \frac{b_1}{b_2} \omega) - (a_1 b_2+a_2
b_1)(\sum_{j=0}^{N-1}\omega^j)$. We note that $Z\neq 0$ and $Z$ is of the form
$\sum_{j}c_j\omega^{j}$ with $c_j\in\ZZ$ and $\sum_{j}c_j=0$ and importantly
that $\arg(Z)=\arg(a'+b'\omega)$. This means we can indeed get within any
$\epsilon$ of our chosen argument $\alpha$ using a judicious choice of
coefficients $c_j$. Hence the resonance points are dense on the $\theta$ axis in
the limit $L\rightarrow\infty$.
 
\begin{figure}
\includegraphics[width=0.45\textwidth]{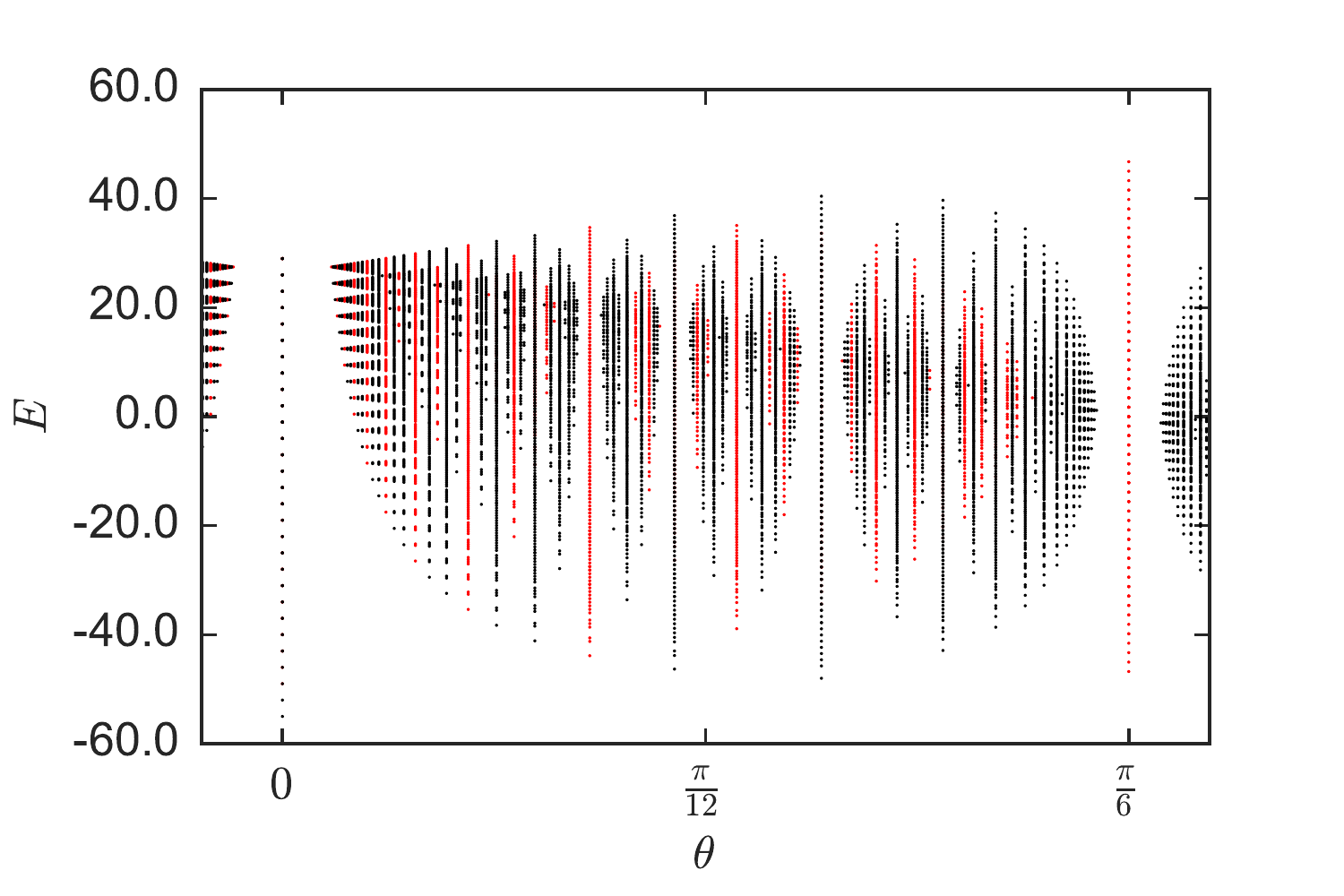}
\caption{All resonance points in the unperturbed spectrum of an $N=3$ system of
  length $L=30$. The red dots indicate resonance points where the total domain wall angle $p$ of the crossing bands is equal. Such resonances do not cause a
  splitting of the topological degeneracy, cf. section \ref{sec:Z3Case}.}
\label{fig:N3L30respoints}
\end{figure}

As an illustration of how the $\theta$ axis is eventually covered by resonance
points, we show all resonance points in an $N=3$ chain of length $L=30$ in
Fig.~\ref{fig:N3L30respoints}. This figure shows that while the resonance
points are eventually dense on the axis, there are notable gaps without any
resonances even at large finite sizes, surrounding the most prominent
resonances. Also higher order resonance points tend to appear at high energy,
meaning that if only states with energy below a given bound are considered,
there may be appreciable regions of the $\theta$-axis where no resonance effects
are observed. 

\section{Perturbation theory}\label{sec:perturbation_theory}

In the last section we explored in detail the appearance and characterisation of
resonance points in the unperturbed spectrum. The reason we are so interested in
 resonance points is that the behaviour of the perturbative series depend
very much on whether one is on- or off-resonance. We show that at the
off-resonant points $q$-sector dependent contributions only occur at an order of
perturbation theory that scales with the length of the system. On the other
hand, when we perform the perturbative expansions at resonant points (where
bands of different domain-wall type and number cross), we typically see that
there are $q$-sector dependent terms which split the degeneracy at an order that
does not scale with $L$. In this section we examine these pertubative series at
both resonant and off-resonant points.

Our approach is to employ the Raleigh-Schrodinger degenerate perturbative
methods pioneered by Kato and Bloch
\cite{Kato1949,Bloch1958,Bloch1958b,Messiah1962,Lowdin1962}. In these approaches
one generates effective hamiltonians acting within a degenerate band of the
unperturbed system, which can be diagonalised to approximate the true
eigenvalues of the full system. We outline the method in detail in appendix
\ref{appx:perturbation_theory}. Here we simply state the effective hamiltonian
$H^{\text{eff} (3)}_{q}$ for a Bloch expansion in the $Q=q$ sector for a band
$b$, to order $f^3$
\begin{align}
\label{eq:Heff3}
& H^{\text{eff} (3)}_{q}  =  P_{0} (H_J + H^{q}_{f})P_{0} + P_{0} H^{q}_{f} \frac{Q_{0}}{E_{0}-H_{0}} H^{q}_{f} P_{0} + \non \\
  & + P_{0} H^{q}_{f} \frac{Q_{0}}{E_{0}-H_0}H^{q}_{f}\frac{Q_{0}}{E_{0}-H_0} H^{q}_{f} P_{0}  \non \\
&     -  P_{0} H^{q}_{f} \frac{Q_{0}}{(E_{0}-H_0)^2} H^{q}_{f} P_{0} H^{q}_{f} P_{0}.
\end{align}
In this expression, $P_{0}$ is the projector onto the band $b$,
$Q_{0}=1-P_{0}$ is its complement, while $E_{0}$ is the unperturbed
energy of the band, as given by Eq.~(\ref{eqn:unperturbed_energy}).

Recall the perturbing hamiltonian in terms of $\alpha$ operators
(\ref{eq:Halpha_q}) is written as
\begin{equation}\label{eqn:Hf}
 H_{f}^q = - f e^{i\phi}\left( \alpha_1 + \sum_{x=1}^{L-2} \alpha_x^{\dagger}\alpha_{x+1}
+ \omega^q\alpha_{L-1}^{\dagger} \right)+ h.c.
\end{equation}
There are only two terms of this operator that actually depend on $q$, the terms
proportional to $\alpha^{\phantom{\dagger}}_{L-1}$ and $\alpha^{\dagger}_{L-1}$.
Only terms in the perturbative expansion which contain these terms can lead to an
energy splitting between $q$-sectors. It is also important to note that both
these terms break total domain-wall angle, defined in ($\ref{eqn:DWParityOp}$).

\subsection{Off resonant $\theta$}
\label{sec:offres}

The main argument we make is that, in the absence of resonance points, the
degeneracy between energy eigenvalues for different q-sectors is exact to an
order of the perturbing parameter that depends on the length of the wire. This is 
provided of course the respective perturbative expansions converge, which is
only the case for sufficiently small perturbing parameter $f$ and sufficiently
far away from resonant points.

Our strategy for the off-resonant scenario is as follows. We show explicitly
that for all bands and all values of N there can be no q-dependence
in perturbative corrections up to the third order. We do this explicitly in this
section up to second order and
discuss generally why this is the case for third order, with additional details
provided in appendix \ref{appx:3rdorder}. For higher orders we have been unable to show
explicitly the lack of q-dependence but argue why this should also be the case.
We also provide numerical evidence that there is no such q-dependence up to
eighth order for some finite systems, and show how these results agree to
machine precision with exact diagonalisation calculations. For brevity some of
the more technical aspects of these calculations have been moved to the
appendix.

\subsubsection{First order corrections}
In the domain-wall picture it is very easy to show that there can be no
$q$-dependent first order corrections for any band $b$. According to
Eq.~(\ref{eq:Heff3}), to get the energy levels of the band to first order in
$f$, we must diagonalise the operator
\[
H^{\text{eff} (1)}_{q}=P_{0}(H_J+H^{q}_{f})P_{0}.
\]  
However, it is not possible to connect two states from band $b$ via either
of the two $q$-dependent terms appearing in $H^{q}_{f_e}$. These terms change the
total domain-wall angle and thus connect to states in different bands,
orthogonal to the original band $b$.
Concretely, if  $\ket{i}$ and $\ket{j}$ are two states in band $b$, then
\[
 \bra{i}  \alpha^{\dagger}_{L-1}
  \ket{j} =  \bra{i} \alpha_{L-1} \ket{j} = 0
\]
Hence all potentially $q$-dependent matrix elements of $H^{\text{eff}(1)}_{q}$ are
actually zero, meaning $H^{\text{eff}(1)}_{q}$ is independent of $q$ and the
eigenvalues of $H$ are independent of $q$ to first order in $f$.

\subsubsection{Second order corrections}\label{sec:2ndorder}

We now show that there is no $q$-dependence in the energy to second order.
According to Eq.~(\ref{eq:Heff3}), to obtain the energy levels of any band $b$
to second order in $f$, we must diagonalise the operator
\[
H^{\text{eff} (2)}_{q}=H^{\text{eff} (1)}_{q}+ P_{0} H^{q}_{f} \frac{Q_{0}}{E_{0}-H_0} H^{q}_{f} P_{0}
\]  
We already know that $H^{\text{eff} (1)}_{q}$ does not depend on $q$, so we can
concentrate on potentially $q$-dependent matrix elements of the second order
part of $H^{\text{eff}(2)}_{q}$. Consider the matrix elements between two states
$\ket{i}=\ket{\textbf{d}^{(i)}}_q^{D}$ and
$\ket{j}=\ket{\textbf{d}^{(j)}}_q^{D}$, in band $b$. We can see from the
fact that the total domain-wall angle of both states is the same, that all
$q$-dependent contributions to the matrix elements of $H^{\text{eff} (2)}_{q}$
connecting these states must involve either
$\alpha^{\dagger}_{L-1}$ and $\alpha^{\phantom{\dagger}}_1$,
$\alpha^{\phantom{\dagger}}_{L-1}$ and $\alpha^{\dagger}_1$ or potentially
$\alpha^{\dagger}_{L-1}$ and $\alpha^{\phantom{\dagger}}_{L-1}$. The
contributions involving both $\alpha^{\dagger}_{L-1}$ and
$\alpha^{\phantom{\dagger}}_{L-1}$ are independent of $q$ since the accompanying
factors of $\omega^{q}$ and $\bar{\omega}^{q}$ cancel any $q$-dependence. The
contributions from other terms add to zero which we will now show. We show this
for $\alpha_{1}$ and $\alpha_{L-1}^{\dagger}$, but the same is true for
$\alpha_1^{\dagger}$ and $\alpha_{L-1}$ (since this is equivalent to exchanging
$\ket{i}$ and $\ket{j}$).

The contributions involved are
\begin{eqnarray}\label{eqn:2ndorderexpression} 
f^2 e^{2i\phi} \omega^q  \left(  \sum_{\ket{k}\not \in b} \frac{ \bra{i} \alpha_1 
\ket{k}\bra{k} \alpha_{L-1}^{\dagger}\ket{j}}{E_{0}-E_k} \right. \non \\
  \left. + \sum_{\ket{l}\not \in b}\frac{\bra{i} \alpha_{L-1}^{\dagger} 
\ket{l}\bra{l}\alpha_1 \ket{j}}{E_{0}-E_l} \right)
\end{eqnarray}
where $\ket{k}=\ket{\textbf{d}^{(k)}}_q^{D}$ and
$\ket{l}=\ket{\textbf{d}^{(l)}}_q^{D}$ run through the eigenstates of $H_0$, which do
not lie in band $b$. $E_k$ and $E_l$ are the unperturbed energies of $\ket{k}$
and $\ket{l}$ respectively. Both the sum over $k$ and the sum over $l$ contain
at most a single nonzero term. These nonzero terms occur in pairs where the
energy denominators have equal magnitude but opposite signs, leading to exact
cancellation (since all nonzero matrix elements of $\alpha_x$ are equal to $1$).

In cases where there are nonzero terms, all the bulk domain-walls in states
$\ket{i}, \ket{j}, \ket{k}$ and $\ket{l}$ must be equal. This means
$d^{(i)}_{x}=d^{(j)}_{x}=d_{x}^{(k)} =d_{x}^{(l)}$ for~all $x\in\{2,...,L-2\}$.
For the domain-wall excitations on the ends, the following conditions must hold

\begin{alignat*}{4}
d^{(i)}_{1}  &= d^{(j)}_{1}-1  &&= d^{(k)}_{1}-1  &&= d^{(l)}_1\\
d^{(i)}_{L-1} &=d^{(j)}_{L-1}+1 &&= d^{(k)}_{L-1}  &&= d^{(l)}_{L-1} + 1
\end{alignat*}
It then follows that
\[
E_{k}=E_{0}+\epsilon_{d^{(i)}_{L-1}}-\epsilon_{d^{(j)}_{L-1}}
\]
and 
\[
E_{l}=E_{0}-\epsilon_{d^{(i)}_{L-1}}+\epsilon_{d^{(j)}_{L-1}}
\]
Summing these we see that $E_0 - E_{k} = -(E_0 - E_{l})$ and the two nonzero
matrix elements between $\bra{i}$ and $\ket{j}$ cancel.

\subsubsection{Third  and higher order corrections}

As one might expect, dealing with third and higher order corrections is 
challenging due to the ever increasing combinations and permutations of terms
that must be considered. In these cases it is instructive to perform the
perturbative expansions numerically. From these calculations, we
observe an energy splitting between $q$-sectors in the third order corrections
at generic values of $\theta$. However this splitting is incrementally reduced
by corrections at higher orders. This leads us to conjecture that in the absence
of resonance points and for sufficiently small $f$, the maximum splitting possible
scales as $f^L$. We proceed by first showing some results from these numerical
calculations and then discussing why we might intuitively expect this behaviour.
In appendix \ref{appx:3rdorder} we show that when using the alternative
perturbative expansion of Soliverez\cite{Soliverez1969} there are no longer any
$q$-dependent third order corrections. We also demonstrate how cancellations
between terms in the perturbative series at third and higher orders can occur.

\begin{figure}
\includegraphics[width=0.40\textwidth]{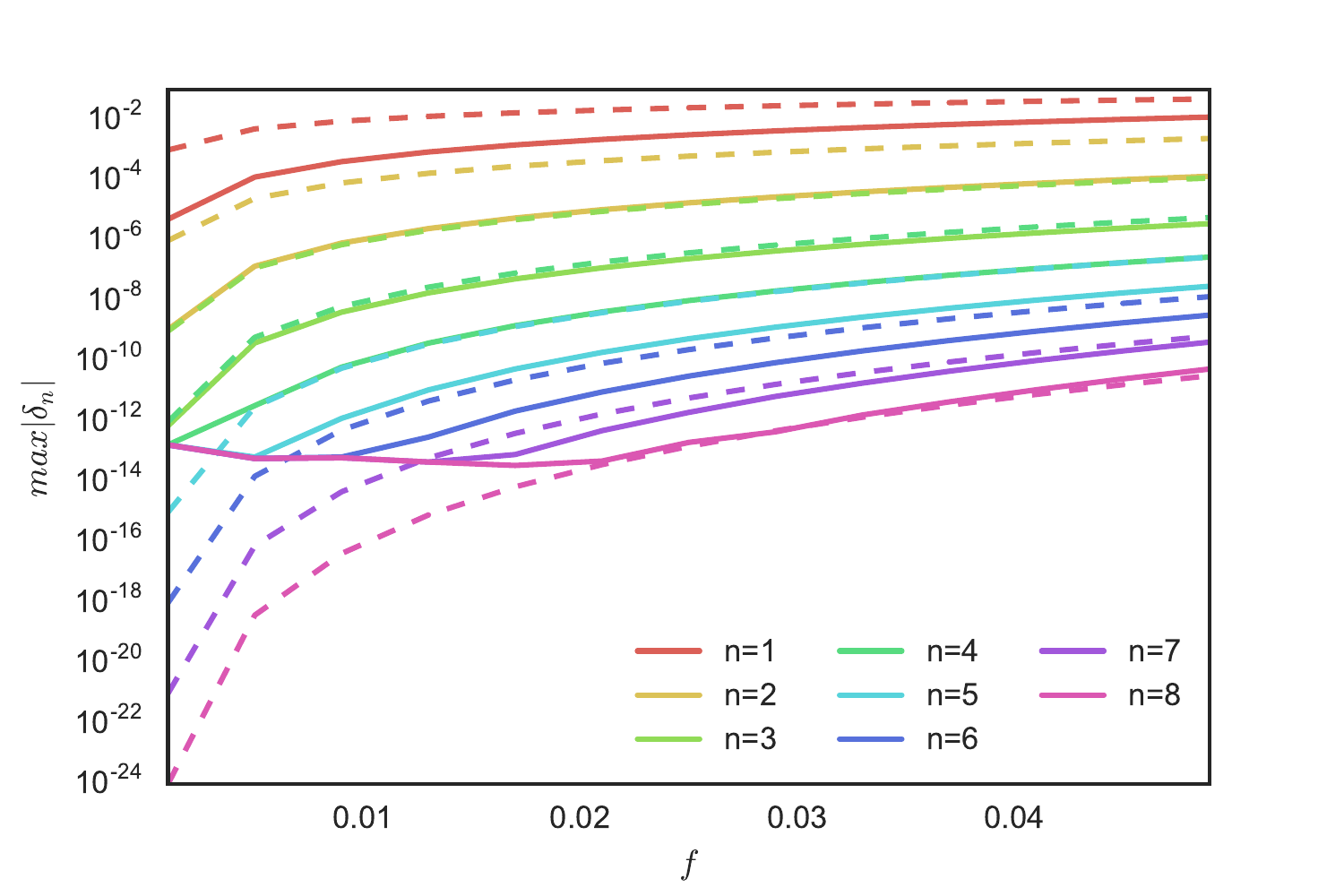}
\caption{The maximum differences between the exact energies (from exact
diagonalisation) and the estimates from $n$th order degenerate perturbation
theory for a range of values of the perturbing parameter $f$. This is for the
first excited band above the ground state for chain of length $L=11$ with $N=3$
and with chiral parameter $\theta=0.3$. The dashed lines show $f^n$ and act as a
guide.}
\label{fig:PTVsExact}
\end{figure}

\begin{figure}
\includegraphics[width=0.40\textwidth]{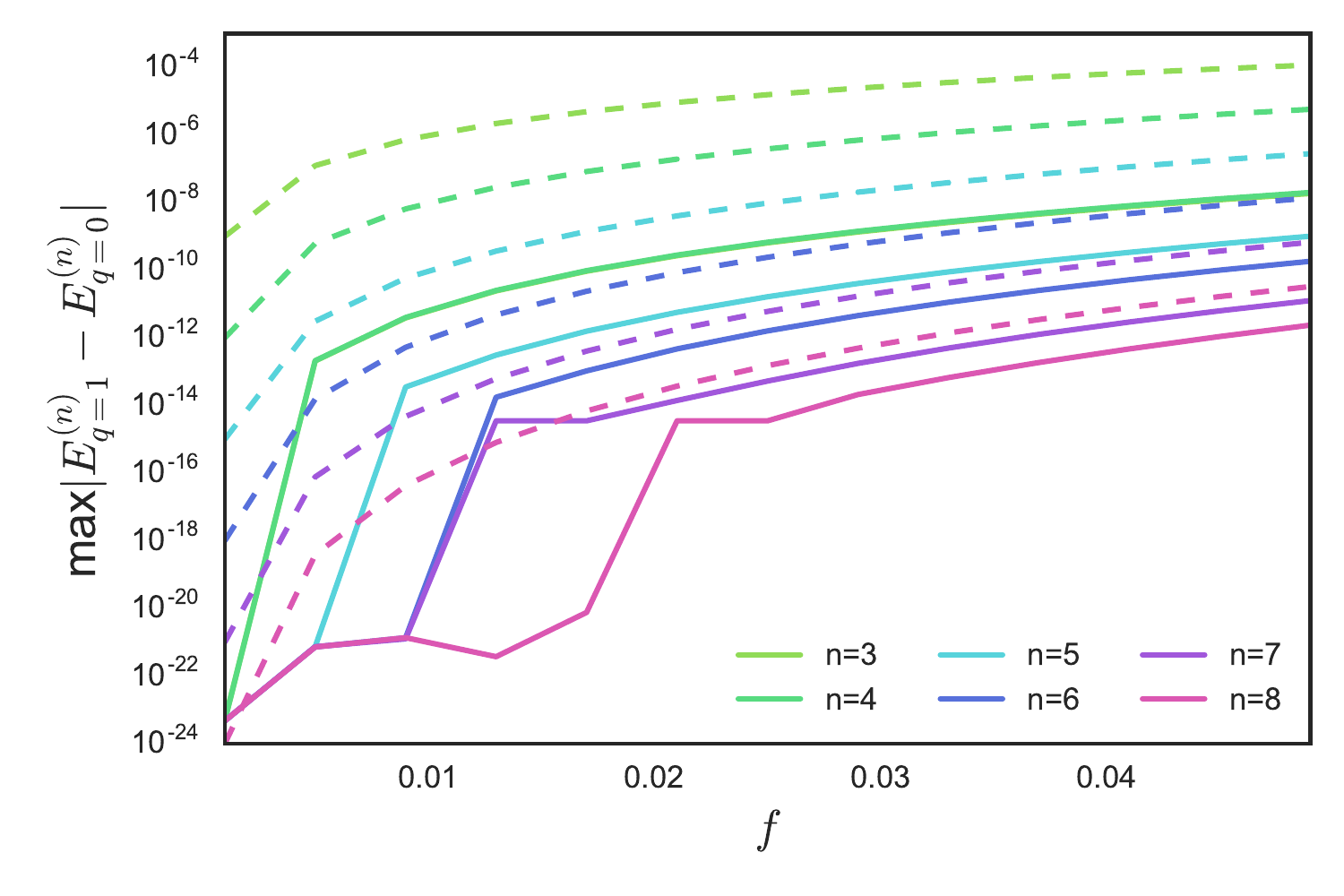}
\caption{The maximum differences between $q=1$ and $q=0$ sectors of the
estimates from $n^{\mathrm{th}}$ order degenerate perturbation theory for a
range of values of the perturbing parameter $f$. This is for the first excited
band above the
ground state, for chain of length $L=11$ with $N=3$ and with chiral parameter
$\theta=0.3$. The dashed lines show $f^n$ and act as a guide.
}
\label{fig:PTBetweenQSectors}
\end{figure}

We use the perturbative expansions of Bloch in our numerical calculations (see
appendix \ref{appx:perturbation_theory} for details). While all expansions are
equivalent in the large order limit, Bloch's expansions contain fewer terms,
making them simpler and more efficient to compute. At modest system size, we can
compare the perturbative results to the exact energy spectra obtained by numerical
diagonalisation. For example, figure \ref{fig:PTVsExact} shows numerical results
for an $N=3$ system at non-resonant values of $\theta$. It is clear that the
results from the numerical perturbation theory calculations agree with the exact
results up to corrections which decrease with the order $n$, whenever they are
not too small to be numerically resolved. In this particular example the
corrections appear to scale as $f^{n+1}$. Similar convergent behaviour is
observed for all other bands. At and in the vicinity of resonance points though it
is necessary to start from a subspace containing all bands which cross at the
resonance point.

Figure \ref{fig:PTBetweenQSectors} shows the maximal difference between the
perturbative energy estimates for states in the $q=0$ and $q=1$ sectors of the
first excited band of the same $N=3$ system. We note that the difference is
nonzero from the third order onwards, but decreases with increasing order
$n\ge3$ and is always small enough to be explained as an error of the
approximation. That is, the energy splitting between the $q=0$ and $q=1$
sectors, calculated from $H^{\text{eff} (n)}$, is less than $f^{n}$. In general
we expect that the maximum difference between $q$-dependent matrix elements of
the effective hamiltonians decay as $f^{n}$. How these matrix elements
effect the energy splitting between $q$-sectors is not completely clear. However
we may argue that these corrections should in general decay faster than $f^{n}$,
since the number of $q$-dependent matrix elements is relatively small and the
dimension of the effective hamiltonians grows with $L$. This behaviour is
observed for all bands as long as $\theta$ is not resonant and $f$ is
sufficiently small. For short chains, when $n \ge L$ a non-vanishing splitting
between $q$-sectors can be observed which is always less than $f^{L}$.

Showing this behaviour directly in the degenerate perturbation series for all
cases is a formidable task. Appendix \ref{appx:3rdorder} shows some specific
examples and arguments in this direction. However we can argue intuitively why
we expect this to be true. For a process to lead to $q$-dependent energy
splitting it must include at least one $q$-dependent term acting on the end of
the chain. Since both these terms change the total domain-wall angle, such a process
must also include a term acting at the start of the chain to connect back to the
original band. This means that at orders lower than $L$ any process that
contributes will contain disconnected sets of operators acting on opposite ends
of the chain. We may intuitively expect that contributions to the energy arising
from operators acting on spatially disconnected parts of the chain should
vanish. General arguments in this direction can be made starting from extensive
energy scaling: If a system consists of independent subsystems the total energy
should equal the sum of the energies of the subsystems and one would expect
cross terms in the perturbation series to vanish, or at least the energy per
site due to such terms should vanish in the thermodynamic limit
$L\rightarrow\infty$ (see e.g. Ref.~\onlinecite{Shavitt2009}). These
considerations are closely related to the existence of linked-cluster-theorems
for degenerate systems where non-local terms corresponding to un-linked diagrams
cancel\cite{Lowdin1951,Brandow1967,Tang2013}. In section
\ref{sec:no_disconnected_pt} we provide numerical evidence that this is indeed
the case by excluding disconnected terms from the pertubative expansions.

Further general arguments that point in this direction can be made based on the
work of Fendley in Ref.~\onlinecite{Fendley2012}. Here for the $\ZZ_3$ model
at generic values of $\theta$ an asymptotic expansion of a parafermionic
zero-mode operator in powers of $f$ (or $f/J$) is well defined.\footnote{Here,
``generic values'' refers to a condition requiring the invertibility of a set of
$\theta$-dependent matrices which appear in the construction of terms in the
expansion. We expect that this is equivalent to the requirement that $\theta$ is
not a resonance point.} For finite systems, the arguments given by Fendley
should indeed imply that the splitting between $q$-sectors vanishes to order
$f^{L}$ at generic $\theta$ and for small enough values of $f$.

\subsubsection{Excluding disconnected terms}
\label{sec:no_disconnected_pt}
It is possible to perform the perturbative expansions numerically in such a way
that processes with operators acting on both ends are explicitly excluded. While
this is not a well defined pertubative expansion, we find that amazingly it
agrees exactly with the exact results up to an error which decreases with the
order of the perturbative expansion $n$, exactly as with the regular
perturbative expansion (figure \ref{fig:PTVsExact}). By construction there is no
$q$-dependence appearing at any order. Of course since this method cannot
account for processes containing operators acting on both ends, we cannot
observe the splitting due to processes connecting both ends of the system. This
is not an issue though when the order $n$ is less than the system size $L$.

The method proceeds by calculating three different effective hamiltonians using
different modifications to the perturbing hamiltonian. We label these effective
hamiltonians $H^{\text{left} (n)}_{q}$, $H^{\text{right} (n)}_{q}$ and
$H^{\text{neither} (n)}_{q}$ where the modifications consist of excluding terms
acting only on the leftmost site, only on the rightmost site and only on the
leftmost or the rightmost sites respectively. Once we have these effective
hamiltonians we combine them to get the final effective hamiltonian as

\begin{equation}
  H^{\text{local} (n)}_{q} = H^{\text{left} (n)}_{q} +  H^{\text{right} (n)}_{q} -  H^{\text{neither} (n)}_{q}
\label{eqn:H_eff_local}
\end{equation}

\subsection{Resonant $\theta$}\label{sec:resonant}
We now turn our attention to the resonance points themselves and see that the
behaviour here is qualitatively different to that found at off-resonant points.
At resonant points, bands of states with different domain-wall configurations
become degenerate making it possible for local processes to
split the degeneracy between $q$-sectors. We begin by looking at first order
processes, particularly those connecting the two lowest lying bands at
$\theta=0$ for the $N=3$ case. Here we write down explicit expressions for the
matrix elements of the effective hamiltonian. We can then diagonalise these
effective hamiltonians analytically to get expressions for the energy splitting
between $q$-sectors. We then look at other higher order resonance points and
using numerical perturbative expansions we pick out the order at which the
splitting occurs. We find that the splittings observed are consistent with the
exact diagonalisation results discussed in section \ref{sec:ED} and with the
characterisation of resonance points given in section
\ref{sec:mapping_resonances}. We then show that not all resonance points lead to
an energy splitting between $q$-sectors. Particular examples of this are the
resonance points at $\theta=0$ for $N=4$ and at $\theta=\frac{\pi}{6}$ for
$N=3$. Both of these are discussed more in section \ref{sec:antires-hidden}.
Note that for convenience, we set $\phi=0$ for most of the rest of this section.

\begin{figure*}
\includegraphics[width=0.80\textwidth]{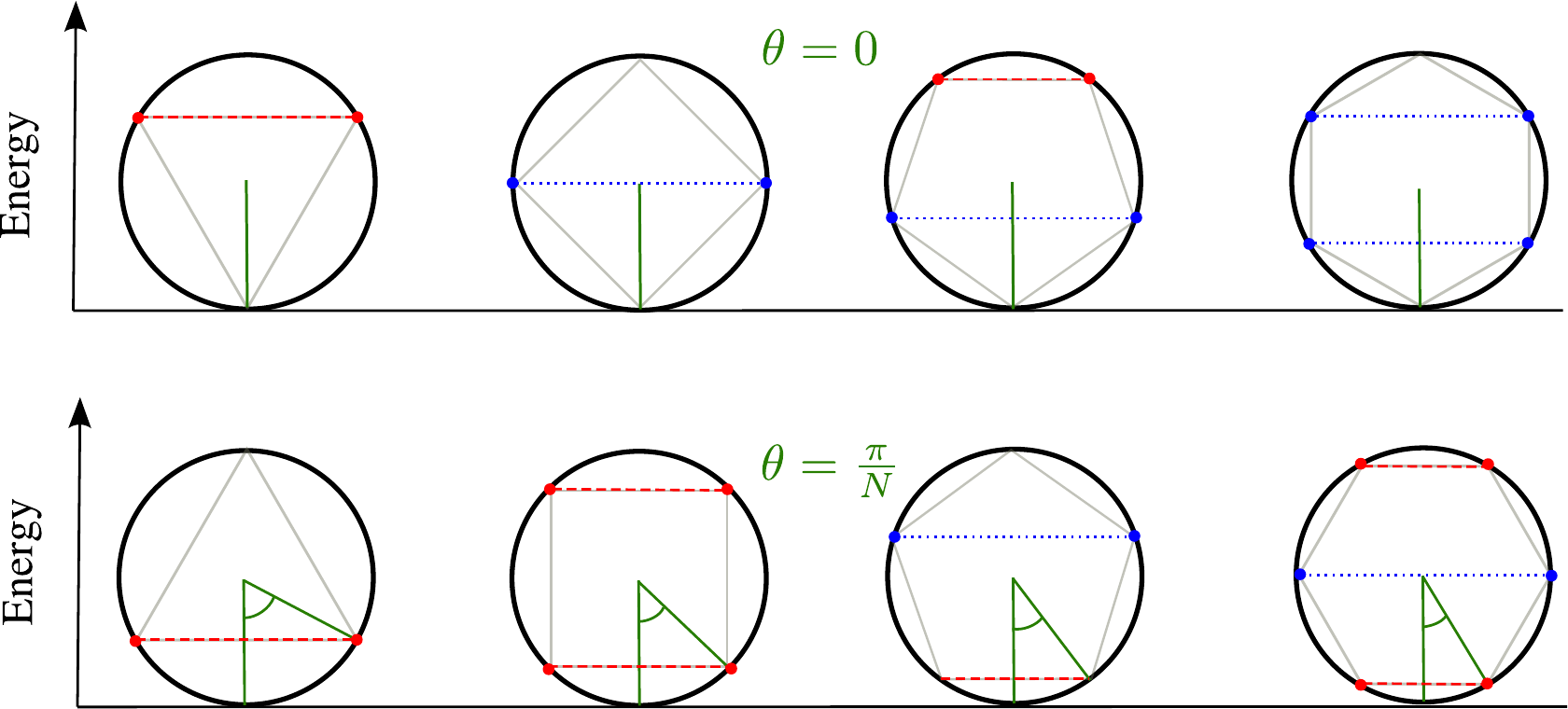}
\caption{Schematic showing simple first-order (red), second-order (blue), and
third-order (purple) resonances in the $N=3,4,5$ and $6$ models for $\theta=0$ and
$\pi/N$. Simple resonances occur when energy levels of different types of domain
walls are the same. They are indicated by the dashed horizontal lines. The order
of the resonance is related to how many step along the circle one must take to
reach the point on the other side of the circle at the same energy. Note that in
the second order resonance for $N=4$ and $\theta=0$ there are two ways to connect
to the other side in two jumps. One through a positive energy state and the
other through a negative energy state.
}
\label{fig:Resonances}
\end{figure*}

\subsubsection{First order resonance points}
We first discuss resonance points where the splitting between $q$-sectors is
first order in $f$. This can only happen when states from two bands that
cross can be connected directly by the $q$-dependent terms in $H^{q}_{f}$,
i.e.~by $f\omega^{q}\alpha^{\dagger}_{L-1}$ and its conjugate. Since these terms
can only change a single domain-wall, it is necessary that there is a
degeneracy between consecutive single domain-wall energies, $\epsilon_{d}=\epsilon_{d \pm 1}$
for some $d$. In Fig.~\ref{fig:Resonances}, we illustrate where degeneracies of
domain-wall energies occur for systems at
$\theta=0$ and $\theta=\frac{\pi}{N}$ for $3\le N\le 6$. We note that
degeneracies between consecutive domain-wall types occur for all $N$ at
$\theta=\frac{\pi}{N}$, whereas at $\theta=0$ we see that adjacent
degenerate wall types occur only for $N$ odd. More generally making use of our
earlier characterisation of resonance points in section
\ref{sec:mapping_resonances}, we observe that first order processes can only
take place at resonance points where $n_{\vec{c}}=2$. There is an additional
constraint that the two non-zero elements of $\vec{c}$ must appear
consecutively, where we consider $\vec{c}$ to be periodic. From plots
\ref{fig:N3L8Map} and \ref{fig:N4L8Map} we see that these only appear when
$\theta$ is a multiple of $\frac{\pi}{N}$.

We now look in detail at the resonance point between the first and second
excited bands of the $N=3$ model at $\theta=0$. These are the bands just above the
ground state in figure \ref{fig:N234Spectrum}. Each of these bands contains a
single domain-wall excitation.  

A useful shorthand for what follows will be to employ a labelling of the states
where only domain-walls with $d\neq 0$ and their locations are indicated,
e.g.
\begin{eqnarray}
\ket{1_x}_q &=& \alpha_x^{\dagger\;1} \ket{\emptyset}_q \non \\
\ket{2_x}_q &=& \alpha_x^{\dagger\;2} \ket{\emptyset}_q \non \\
\ket{1_{x_1} 3_{x_2}}_q &=& \alpha_{x_1}^{\dagger\;1} \alpha_{x_2}^{\dagger\;3} \ket{\emptyset}_q 
\label{eqn:band_shorthand}
\end{eqnarray}

Using this shorthand, we label states in these bands $\ket{1_x}_q
=\alpha_x^{\dagger \; 1} \ket{\emptyset}_q$ and $\ket{2_x} =\alpha_x^{\dagger\;
2} \ket{\emptyset}_q$. The only $q$-dependent first order processes that connect these states
occur at the end. The relevant matrix elements are (setting $\phi=0$)
\begin{eqnarray}
~_q\bra{2_{L-1}} \omega^q  \alpha^{\dagger}_{L-1} \ket{1_{L-1}}_q  &=\omega^q
                                                                     \non \\ 
~_q\bra{1_{L-1}} \omega^{-q} \alpha^{\phantom \dagger}_{L-1} \ket{1_{L-1}}_q
                                          &=\omega^{-q} \non
\end{eqnarray}
Ordering the basis elements as: \begin{equation}
\ket{1_{L-1}}_q,\ket{1_{L-2}}_q, ... ,\ket{1_{1}}_q, \ket{2_1}_q, ...,
\ket{2_{L-2}}_q, \ket{2_{L-1}}_q \non \end{equation} we see that the effective
hamiltonian for the two bands can be written as a $2L-2$ Toeplitz matrix:
\begin{equation}
H^{(1)}_f= -f  \left[ \begin{array}{cccccc}
0 & 1  & 0  & \hdots & 0&  \omega^{-q} \\
1 &  0 & 1  & \hdots & 0& 0 \\
 0 & 1 & & &:  &: \\
: & :  &   &   &1 & 0\\
 0 & 0 & \hdots &1  &0 & 1 \\
 \omega^{q} & 0  & \hdots   & 0 & 1 &0 \end{array} \right]
\label{eqn:H1_single}
\end{equation}
which can be diagonalised with a generalised
Discrete-Fourier-Transform:\footnote{We give the results below for general $N$
because these are also the energy splittings for resonances between the
lowest bands at $\theta=\frac{\pi}{N}$, for arbitrary $N$.}
\begin{equation}
F_{mn}= \frac{1}{\sqrt{L'}} e^{i \frac{2 \pi}{L'} (m-1+\frac{q}{N}) \times (n-1+\frac{q}{N}) }
\end{equation}
where $L'=2L-2$. This gives a spectrum of the form
\begin{equation}
E(k)_q= - 2 f \cos \left(\frac{2 \pi}{L'} (k+ \frac{q}{N}) \right)
\end{equation}
where $k \in [1, ..., L']$. The difference in energies between two states $
\Delta E= E(k_1)_{q_1} - E(k_2)_{q_2} $ is therefore
\begin{eqnarray}
\label{eq:DEscaling}
\Delta E &= - 4 f &\sin \left( \frac{\pi}{L'}(k_1 - k_2 + \frac{q_1-q_2}{N}) \right) \times \non\\ 
         && \sin \left(  \frac{\pi}{L'} (k_1 + k_2 +\frac{q_1+q_2}{N}) \right).
\end{eqnarray}
The relevant comparison to make for the $q$-dependent splitting is between
states where $k_1=\pm k_2$, as these states would have the same energy when
$\Delta q = q_1 - q_2$ is taken to zero.
In these cases, for large $L$ we find that $\Delta E \propto \frac{f}{L^2}$ if $k_1\ll L$
while $\Delta E \propto \frac{f}{L}$ if $k_1\approx L/2$. We also note that
there is no splitting if $\Delta q=N$. In this case the states
$\ket{k_1}_{q_1}$ and $\ket{-k_1}_{q_2}$ remain exactly degenerate as a
consequence of the dihedral symmetry (which is present since we set $\phi=0$).
At $N=3$, this degeneracy is observed when $q_1=1, q_2=2$. Of course it is also
clear that there can be no splitting between sectors in this case as the
effective matrix (\ref{eqn:H1_single}) for $q=1$ is the Hermitian conjugate of
the effective matrix for $q=2$.

For higher energy bands at $\theta=0$, there are many more bands crossing and
each of these bands has a larger numbers of states (we may for example consider
bands with states $\ket{1_x,1_y}$, $\ket{1_x, 2_y}$, $\ket{2_x, 2_y}$). Many of
these bands can still be connected via first order processes like the ones we
just demonstrated.
The result is that the splitting between $q$-sectors here also scales with $f$.
The factor will in general be smaller than in \eqref{eq:DEscaling}, since the
effective matrices will have larger dimension and thus the $q$-dependent matrix
elements have less impact on the eigenvalues.

Similar results for this first order crossing where found by Jermyn et. al. in
ref. \onlinecite{Jermyn2014}, although using their notation the effective
matrices are not written directly in the each $q$-sector and by
projecting to sectors with constant domain-wall number, it is not possible to
observe higher order resonance points at $\theta \ne 0$.

\subsubsection{Higher order resonance points}

Resonance points can also introduce $q$-dependent splitting at higher orders in
$f$. These can occur between bands which differ only at a single domain
wall, e.g.~the bands containing states $\ket{d_x}$ and $\ket{(d+m)_{x}}$ are in
resonance if $\epsilon_{d}(\theta)=\epsilon_{d\pm m}(\theta)$ which can can
potentially cause splitting at order $l=\min(m,N-m)$.
However, most higher order resonances are not of the simple type just described.
Instead they result from combinations of domain-wall energies that become
degenerate. Using the notation for bands that was introduced in section
\ref{sec:mapping_resonances}, we describe a resonance point between two bands
$\vec{a}$ and $\vec{b}$ by a vector $\vec{c} = \vec{a} - \vec{b}$. We then write
$n_{\vec{c}}$ for the Hamming distance between $\vec{a}$ and $\vec{b}$. If the
bands are in resonance, then the order of perturbation theory at which
$q$-dependent splitting appears, $m_{\mathrm{split}}$, must satisfy 
\begin{equation}
\label{eq:orderbound}
m_{\mathrm{split}}\ge 1+(n_{\vec{c}}-2)/4
\end{equation}
To see this, note that the application of $H_{f_e}^{q}$ to
the band with domain-wall vector $\vec{a}$ changes the Hamming distance of the
state to $\vec{b}$ by $2$, while application of $H_{f_b}^{q}$ changes the
Hamming distance by at most $4$. At least one application of $H_{f_e}^q$ is
necessary to introduce $q$-dependence. Hence to get from $\vec{a}$ to $\vec{b}$
we need to operate with $H_{q}^{f}$ at least $1+(n_{\vec{c}}-2)/4$ times. This
gives a lower bound for the order at which $q$-dependent splitting can occur at
such a resonance point.

\begin{figure}
\includegraphics[width=0.45\textwidth]{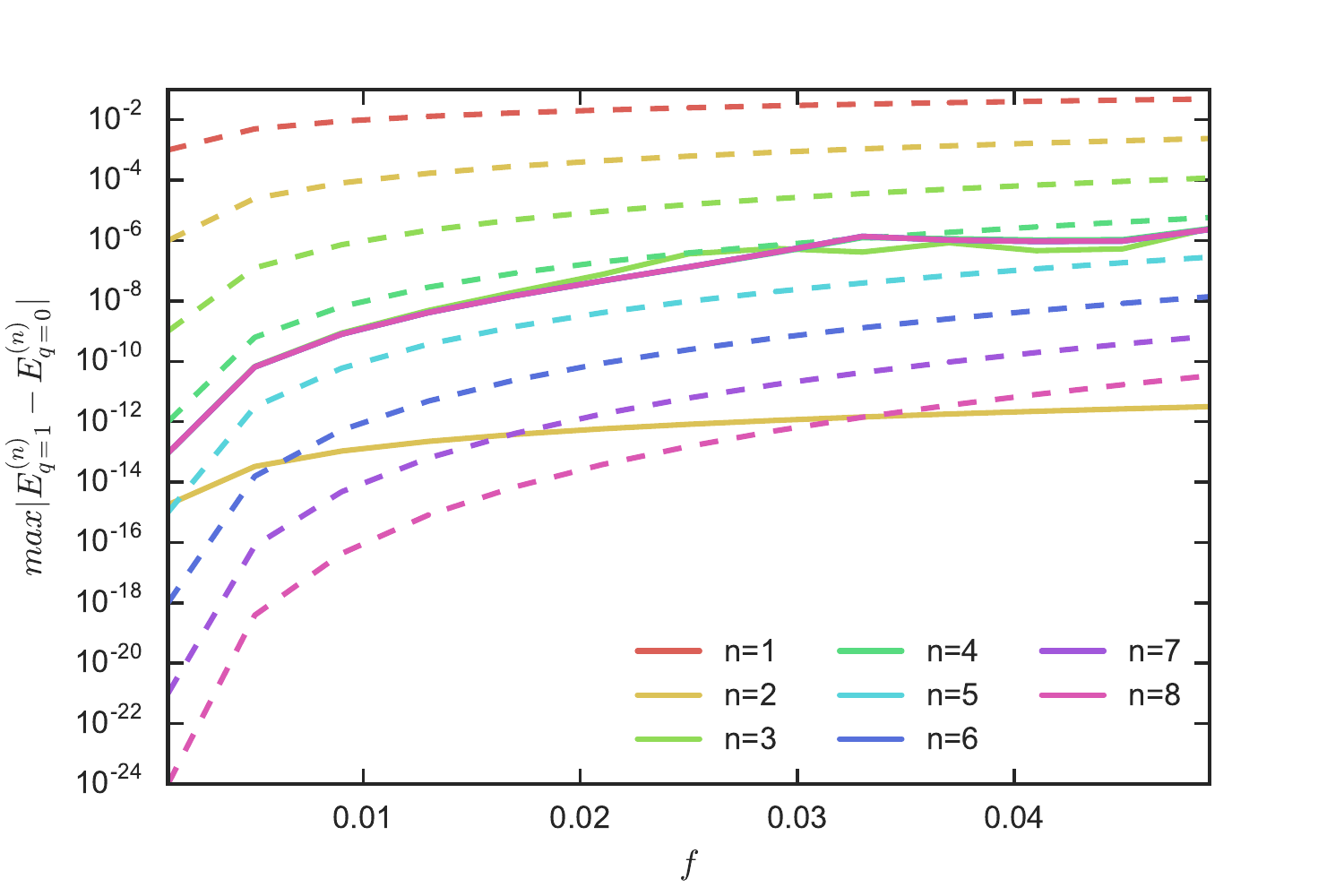}
\caption{The maximum differences between $q=1$ and $q=0$ sectors of the
estimates from $n$th order degenerate perturbation theory for a range of values
of the perturbing parameter $f$. This is for the fifth and sixth excited bands
for a chain of length $L=8$ with $N=3$ and with chiral parameter
$\theta=\arctan(\frac{\sqrt{3}}{5})$. The dashed lines show $f^n$ and act as a
guide. The degenerate space from which we perturb here has dimension 56.} 
\label{fig:PT_bands_5_6_order_3_splitting}
\end{figure}

As a specific example consider the resonance point between the fifth and sixth
excited bands of the $N=3$ model. This resonance is marked in figure
\ref{fig:N3L8Map} and has $n_{\vec{c}}=6$. The lower (upper) band from the point
of view of increasing $\theta$ contains states of type $\ket{1_x,1_y}$
($\ket{2_x 2_y 2_z}$). Using the notation introduced in section
\ref{sec:mapping_resonances}, these are characterised by the tuples
$\vec{a}=(L-3,2,0)$ and $\vec{b}=(L-4,0,3)$ respectively. The domain-wall
energies at this resonance point satisfy $\epsilon_{0}+2\epsilon_{1}=3\epsilon_{2}$.
Using Eq.~(\ref{eqn:phi_at_crossing}), we find the exact $\theta$ for this
crossing is $\theta=\arctan(\frac{\sqrt{3}}{5})$. From the domain-wall content
of these bands, we see that the lowest order on which these can in principle be
connected with $H_f^q$ is the second order. However, the $q$-dependent second
order terms cancel and we find $q$-dependent splitting at third order. Figure
\ref{fig:PT_bands_5_6_order_3_splitting} shows results of numerical perturbative
expansions at this point for a range of $f$ values. We see that there is
splitting in the energy between $q$-sectors appearing at third order. Unlike the
behaviour at the off-resonant points (figure \ref{fig:PTBetweenQSectors}) where
the spltting is reduced at higher orders, here it remains. This agrees with
observations from the exact numerics shown in figure
\ref{fig:ExactEnergySplitting} and discussed in section \ref{sec:ED}.

\section{Full diagonalisation of small systems}
\label{sec:ED}
\begin{figure}
\includegraphics[width=0.40\textwidth]{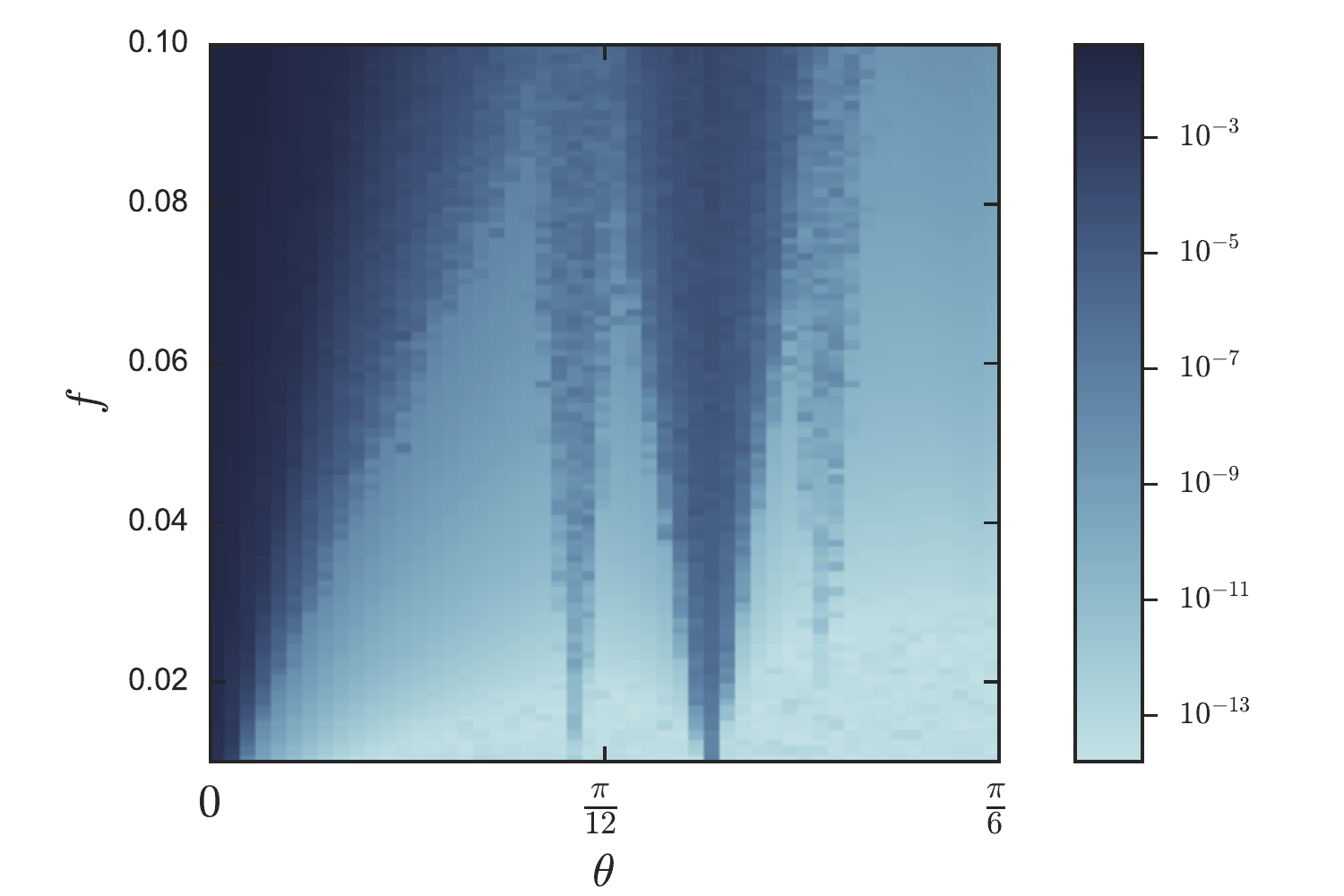}
\caption{Plot of the maximum energy splitting between the $q=0$ and $q=1$ sectors
  for a chain of length $L=9$ with $N=3$. Results obtained using exact
diagonalisation where each data point corresponds to maximum difference between
all 6561 eigenvalues in the $q=0$ and $q=1$ sectors.}
\label{fig:ExactEnergySplitting}
\end{figure}

For small systems we can examine the maximum splitting between $q$-sectors over
a range of values of $\theta$ and $f$ by fully diagonalising the hamiltonian. We
can do this directly in each $q$-sector using the domain-wall picture discussed
in section \ref{sec:domain_wall_picture}. Figure~\ref{fig:ExactEnergySplitting}
shows a plot of the maximum splittings between the $q=0$ and $q=1$ sectors for
an $N=3$ system with $L=9$ (note that the $q=1$ and $q=2$ sectors are exactly
degenerate due to the dihedral symmetry described in section
\ref{sec:dihedral}). In each $q$-sector the energy levels have been ordered and
we have plotted the maximal absolute difference between eigenvalues at the same
position in this ordering.

It is clear from figure \ref{fig:ExactEnergySplitting} that the largest
splitting between sectors occurs around the $\theta=0$ point. This can be
attributed to the first order processes occurring at resonance points here, which
are discussed in detail in section \ref{sec:resonant}. For larger $f$, we see
other branches of splittings emerge. These can be attributed to third, fourth
and fifth order processes occurring at the resonance points indicated in figure
\ref{fig:N3L8Map}. While there are many resonance points at
$\theta=\frac{\pi}{6}$, these do not result in any splitting between the
sectors. This is because the bands crossing here have the same total domain-wall
angle. This is examined in more detail in section \ref{sec:Z3Case}. Away from
resonance points, we see that there is very little splitting (especially at low
$f$). This is consistent with the perturbation theory results for off-resonant
$\theta$ in section~\ref{sec:offres}. Some splitting does appear at larger $f$
even away from resonances, but this can be ascribed to finite size effects which
allow for $q$-dependent perturbative processes at order $f^{L}$.

\section{Anti-resonance and hidden zero-modes in the $\ZZ_3$ and $\ZZ_4$ models}
\label{sec:antires-hidden}
In previous sections we outlined why for $N$-prime, finite chain length $L$,
sufficiently small $f$ and at generic values of the chiral parameter $\theta$ we
observe that the energy splitting between $q$-sectors decays exponentially as
the length of the chain is increased. We were also able to show that the regions
where this $f^L$ splitting is observed become vanishingly small as $L$ gets
larger, and hence there is typically no strong zero-mode in the thermodynamic
limit. We also showed that strong zero-modes do not typically exist anywhere if
$N$ is composite, owing to $q$-dependent terms appearing at very low orders
between bands that are everywhere degenerate in the un-perturbed limit.

In this section we outline counter-examples where these general observations do
not hold. The first case is for the region around $\theta=\frac{\pi}{6}$ of the
$N=3$ model which was previously noted in Ref. \onlinecite{Jermyn2014} as being
the most likely region to support strong zero-modes. We show below that, despite
the presence of prominent resonance points at $\theta=\frac{\pi}{6}$ (see Figure
\ref{fig:N234Spectrum}), these do not result in any $q$-dependent power-law energy
splitting. This fact, coupled to the knowledge that $N$ in this case is prime,
allows us to argue that a region $\delta \theta$ about this point is also
degenerate to order $f^L$, but that the size of this stable region decreases as
$\delta \theta \sim 1/\sqrt{3} L$. 
We go on to show that there are in fact additional values of $\theta$ where the
resonance points do not result in any $q$-dependent power-law splitting. However
these are less prominent and have a smaller surrounding stable region than for
$\theta=\frac{\pi}{6}$.

The second case is the $\theta=\frac{n\pi}{2}$ points of the $N=4$ model. These
$\theta$ values are also prominent resonant points (see Figure
\ref{fig:N234Spectrum}), but are actually exactly solvable; mappable to two
uncoupled Majorana chains. Although the case for strong-modes in this scenario
is non-perturbative, because the $N=4$ system is composite, we actually see that
the degeneracy away from this point is broken at low orders of $f$ and so the
parafermionic strong zero-modes can only exist exactly at the special point i.e
$\delta \theta =0$. We make the case however that there exist hidden
topological zero-modes between some of the q-sectors and write down what these
modes look like in position space using an iterative approach similar to that
employed in Ref. \onlinecite{Fendley2012}.
 
\subsection{The $\ZZ_3$ system in the vicinity of $\theta=\pi/6$}
\label{sec:Z3Case}
\begin{figure}
\includegraphics[width=0.30\textwidth]{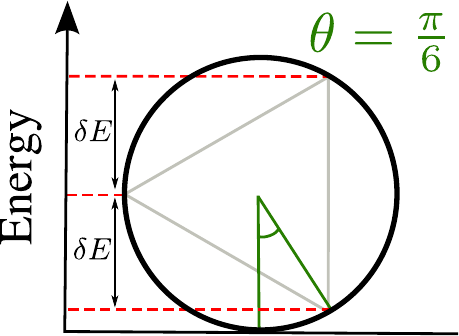}
\caption{Domain-wall excitation energies at $N=3$ and $\theta=\frac{\pi}{6}$ are equally spaced.} 
\label{fig:piover6}
\end{figure}

The $\theta=\frac{\pi}{6}$ point at $N=3$ is unique for odd $N$ in that the
single domain-wall energies at this point are equally spaced, see
Fig.~\ref{fig:piover6}. This property is quite powerful as it is one of the main
ingredients in showing the exact solvability and $N$-fold degeneracy of a class
of $\ZZ_N$ models studied in Ref.~\onlinecite{Gehlen1985}. In particular, the so
called super-integrable point of the $N=3$ model also occurs at
$\theta=\frac{\pi}{6}$, but in addition has $\phi=\frac{\pi}{6}$.

On a perturbative level this precise match-up between domain-wall excitation
energies is also very important because it brings together states (in energy)
that can only be connected by total domain-wall angle (see \eqref{eqn:DWParityOp})
preserving terms. This means that the quadratic bulk terms $H_{f_b}^q$ in
\eqref{eqn:Hf_edge} can quite easily map between states in the two bands (even on
the first order) but that terms that break this total domain-wall angle (those
from $H_{f_e}^q$ from (\ref{eqn:Hf_edge})) must occur in total domain-wall angle
preserving pairs. As a consequence the qualitative behaviour of the PT at this
point is identical to that of the off-resonant PT discussed in section
\ref{sec:offres}.

\begin{figure}
\includegraphics[width=0.45\textwidth]{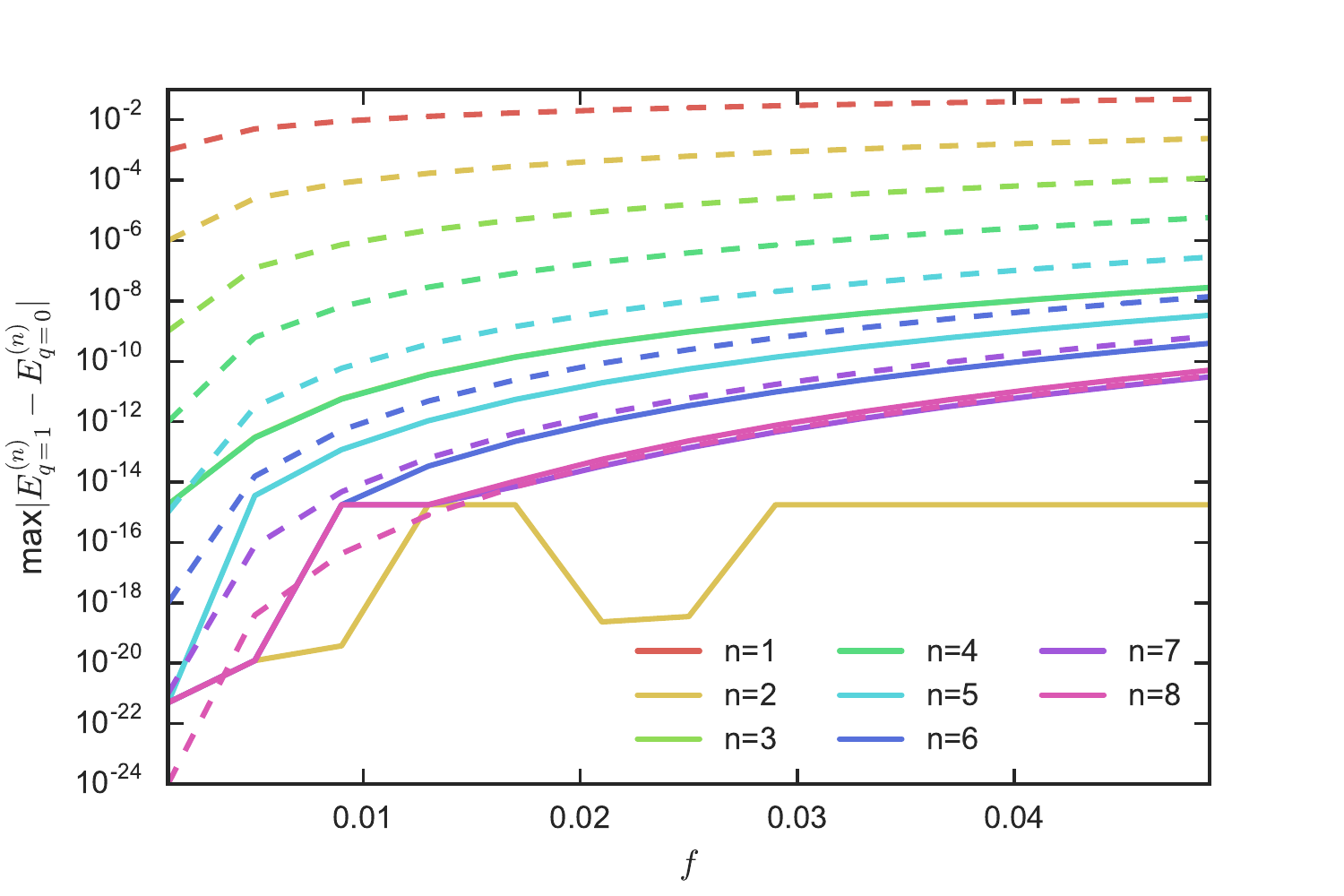}
\caption{The maximum differences between $q=0$ and $q=1$ sectors of the
estimates from $n$th order degenerate perturbation theory for a range of values
of the perturbing parameter $f$. This is for the second and third excited bands
for a chain of length $L=8$ with $N=3$ and with the chiral parameter set to the
prominent resonance point at $\theta=\frac{\pi}{6}$. The dashed lines show $f^n$
and act as a guide. The degenerate space from which we perturb here has
dimension 28.}
\label{fig:PT_N3_bands_2_3_phi_pi_over_6}
\end{figure}

As additional evidence for this claim we note that the effective hamiltonians
generated within the degenerate subspace are $q$-independent up to the third
order, when using Soliverez's \cite{Soliverez1969} symmetrized perturbative
expansion. Further numerical evidence is provided in Figure
\ref{fig:PT_N3_bands_2_3_phi_pi_over_6} where we note that, identical to the
off-resonant cases, the $q$-dependency becomes progressively smaller (in a manner
consistent with a topological degeneracy) as we include more orders in the
expansion. 

For finite $L$, there is a substantial region around the $\theta=\frac{\pi}{6}$
point which can contain exact zero-modes, since the
resonance points which break the degeneracy only approach $\theta=\frac{\pi}{6}$
slowly with increasing $L$, see figures \ref{fig:N3L30respoints} and
\ref{fig:ExactEnergySplitting}. The resonances near $\theta=\frac{\pi}{6}$ also
occur at high energy (in the middle of the spectrum) and at high orders in $f$.
It is therefor likely that even in the thermodynamic limit, there will be many
states, even states with finite energy density, for which the topological
degeneracy is preserved to extremely good approximation. We should also mention
here that the perturbation theory and exact diagonalisation give the same
qualitative results regardless of value chosen for the other chiral parameter
$\phi$. This suggests that strong zero-modes exist all along the
$\theta=\frac{\pi}{6}$ line. Note that their existence is not dependent on the
super-integrability at $\theta=\pi/6$ as this occurs only at the point
$(\theta,\phi)=(\frac{\pi}{6},\frac{\pi}{6})$.

We can find other so-called anti-resonance points where bands with the same
total domain-wall angle cross and thus no $q$-dependent splitting results. This
is illustrated in figure \ref{fig:N3L30respoints} where the anti-resonance points are shown in
red. These points though are not as prominent as the point at $\theta=\frac{\pi}{6}$
because: 1) the bands are not as well separated and thus perturbative
arguments are only valid for very small $f$ and 2) the distance to surrounding
resonance points is much smaller meaning the stable region in the vicinity of
these points is also much smaller. 

To find the locations of the anti-resonances analyticallly, we first write the
single domain wall energies $\epsilon_i$ for $i>0$ in the form
$\epsilon_i=\epsilon_0+m_i\delta$ where $\{m_i\}$ are integers and $\delta$ is a
real number. Using equation (\ref{eqn:unperturbed_energy}) we find that at a
resonance point described by vector $\vec{c}$ we have $\sum_ic_i\epsilon_i=0$
which implies that
\begin{equation}
  \sum_{i>0}c_im_i=0
  \label{eqn:cnst1}
\end{equation}
The total domain-wall angle ($\ref{eqn:DWParityOp}$) of a band described by
vector $\vec{a}$ is $p_{\vec{a}}=\sum_i ia_i \Mod{N}$. At resonance points where
both bands have the same domain-wall angle
\begin{equation}
  \sum_i ic_i \Mod{N}=0
  \label{eqn:cnst2}
\end{equation}
must hold. While (\ref{eqn:cnst1}) and (\ref{eqn:cnst2}) are valid for all $N$
when $N=3$, they lead to the straightforward condition that $m_1+ m_2 = 0
\Mod{3}$. Using the expression for the single domain-wall energies from
(\ref{eqn:DWH0Energy}) we find that for a given $\{m_i\}$ satisfying this
constraint the resonance point can be found at $\theta=\arctan\left(
\frac{\sqrt{3}(m_1-m_2)}{m_1+m_2} \right)$ (assuming the system is sufficiently
large).

For example, the simplest solution $(m_1,m_2) = (2,1)$ corresponds to the
$\theta=\frac{\pi}{6}$ point. Another solution $(m_1,m_2)=(5,4)$ gives
$\theta=\arctan\left( \frac{1}{3\sqrt{3}} \right)$ which corresponds to the line
of $n_{\vec{c}}=10$ resonance points on the left side of figure
\ref{fig:N3L8Map}. From the exact numerics shown in figure
\ref{fig:ExactEnergySplitting} we see that there is indeed no apparent splitting
in the vicinity of this point, while there is for the other line of
$n_{\vec{c}}=10$ resonance points on the right side of the figure.

\subsection{The $\ZZ_4$ system}
\label{sec:Z4Case}

\begin{figure}[h]
\includegraphics[width=0.45\textwidth]{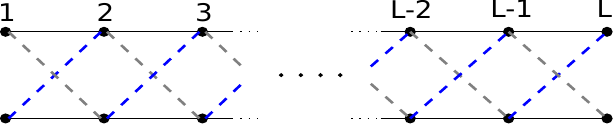}
\caption{ The $\mathbb{Z}_4$ model decouples into two decoupled
  spin-$\frac{1}{2}$ chains when $\theta = \frac{n\pi}{2}$ and $\phi = m \pi$. 
  For $\theta = n \pi$ these are transverse Ising chains defined on the legs
  (solid black lines), whereas for $\theta = n \pi + \frac{\pi}{2}$ these are
  zig-zag chains which opposite signs, depicted here with blue and grey dashed
lines.}
\label{fig:Z4mode}
\end{figure}

Let us first show how the $N=4$ model can be written as a spin-ladder that
decouples into two spin-$\frac{1}{2}$ chains when $\theta=\frac{n\pi}{2}$ and
$\phi = m \pi$ for integer $n,m$. At these special points the resulting model
can be solved exactly (for all $f$). To see how this comes about, we first
introduce the local unitary transformation $\xi$ which permutes the third and
fourth clock states. We write this as an operator on
$\CC^{4}=\CC_u^2\otimes\CC_d^2$ as follows
\[
\xi=\frac{1}{2} (I+\sigma_{d}^{x}-\sigma_{u}^{z} \sigma_{d}^{x}+\sigma_{u}^{z})
\] 
where we have introduced two copies of the Pauli matrices (up and down). The
operator $\xi$ satisfies the following properties,
\begin{equation}
\label{eq::ts}
\begin{split}
&\xi\sigma\xi^\dagger=\left( \frac{1+i}{2}\right)\sigma^z_u+ \left( \frac{1-i}{2}\right)\sigma^z_d\\
&\xi\tau\xi^\dagger=\frac{1}{2}(\sigma^x_u+\sigma^x_d)+\frac{i}{2}(\sigma^z_u\sigma^y_d-\sigma^y_u\sigma^z_d)\\
\end{split}
\end{equation}
Thus after applying the unitary operator $\Xi=\prod_{i=1}^{L}\xi_{i}$, the
$N=4$ hamiltonian takes the following form,
\begin{eqnarray}
\Xi \left(H_J +H_f\right) \Xi^\dagger &=& 
 -J \cos (\theta)  \sum_{i=1}^{L-1} \left( \sigma^z_{i,u}
                                          \sigma^z_{i+1,u}+\sigma^z_{i,d}
                                          \sigma^z_{i+1,d} \right) \nonumber  \\
&-& J\sin(\theta) \sum_{i=1}^{L-1} \left( \sigma^z_{i,u} \sigma^z_{i+1,d}-\sigma^z_{i+1,u}
    \sigma^z_{i,d}\right)\nonumber \\
&-&f \cos(\phi)\sum_{i=1}^L \left( \sigma^x_{i,u}  + \sigma^x_{i,d} \right)\nonumber \\
&-&f \sin(\phi)\sum_{i=1}^L \left( \sigma^{y}_{i,u}\sigma^{z}_{i,d} - \sigma^{z}_{i,u}\sigma^{y}_{i,d}\right)
\end{eqnarray}
Setting $\theta = \frac{n\pi}{2} $ and $\phi= m \pi$, we see that
many of the terms vanish and we end up with two decoupled chains. For $\theta =
n \pi$, these are transverse Ising chains on the legs of the ladder, whereas
for $\theta = n \pi + \frac{\pi}{2}$ they are zig-zag chains with alternating
ferromagnetic and anti-ferromagnetic couplings. This structure is shown in
figure \ref{fig:Z4mode}.

When the system can be written as decoupled spin-$\frac{1}{2}$ chains
(for $\theta=\frac{n\pi}{2}$ and $\phi=m \pi$), there is an exact four fold
degeneracy and we expect there to be exact parafermionic zero-modes. Here we
explore the case where $\theta=0$ and write expressions for these in terms of
the fermionic zero-modes of the decoupled chains. Analogous constructions are possible
for other special points. Starting with the zero-modes for the two Ising chains
the resulting parafermionic zero-modes:
\begin{equation}
\begin{split}
\alpha_L&=\frac{1}{\sqrt{2}}(e^{i\frac{\pi}{4}}\alpha_L^u-e^{-i\frac{\pi}{4}}\alpha_L^d)\\
\alpha_R&=\frac{1}{\sqrt{2}}(\alpha_{R}^{u}\,Q^2-i\alpha_{R}^{d})T
\end{split}
\end{equation} 
where the operators $\alpha^{u/d}_{L/R}$ are Majorana zero-modes acting at
the left and right hand edge of the upper and lower Ising chains. The
operator
\begin{equation}
T=\frac{1}{2}\prod_{i=1}^L(I_i+\sigma_{i,u}^z\sigma_{i,d}^z+\sigma_{i,u}^x\sigma_{i,d}^x+\sigma_{i,u}^y\sigma_{i,d}^y)
\end{equation}
exchanges the states between upper and lower chains and satisfies
\begin{equation}
T\alpha^{u}_{L/R}=\alpha^{d}_{L/R}T 
\end{equation}

Using the anti-commutation relations of the fermionic zero-modes (note that
zero-modes on different chains commute), and the fact that 
\begin{eqnarray}
Q&=&TP_d=P_uT \Longrightarrow Q^2=P_uP_d \non \\
Q^2T&=&TQ^2
\end{eqnarray}
where $P_d=\prod_{i}\sigma_{i,d}^{x}$ and $P_u=\prod_{i}\sigma_{i,u}^{x}$ are
the $\ZZ_2$ symmetry operators (or fermionic parity operators) of the upper and
lower chains, one may check directly that the $\alpha_{L}$ and $\alpha_{R}$
given above satisfy the correct parafermionic relations (\ref{eq:paraf_alg}), in
this case $\alpha_{L}^{4}=\alpha_{R}^{4}=1$ and
$\alpha_{L}\alpha_{R}=-i\alpha_{R}\alpha_{L}$.

\subsubsection{Stability and hidden zero-modes}

\begin{figure}
\includegraphics[width=0.45\textwidth]{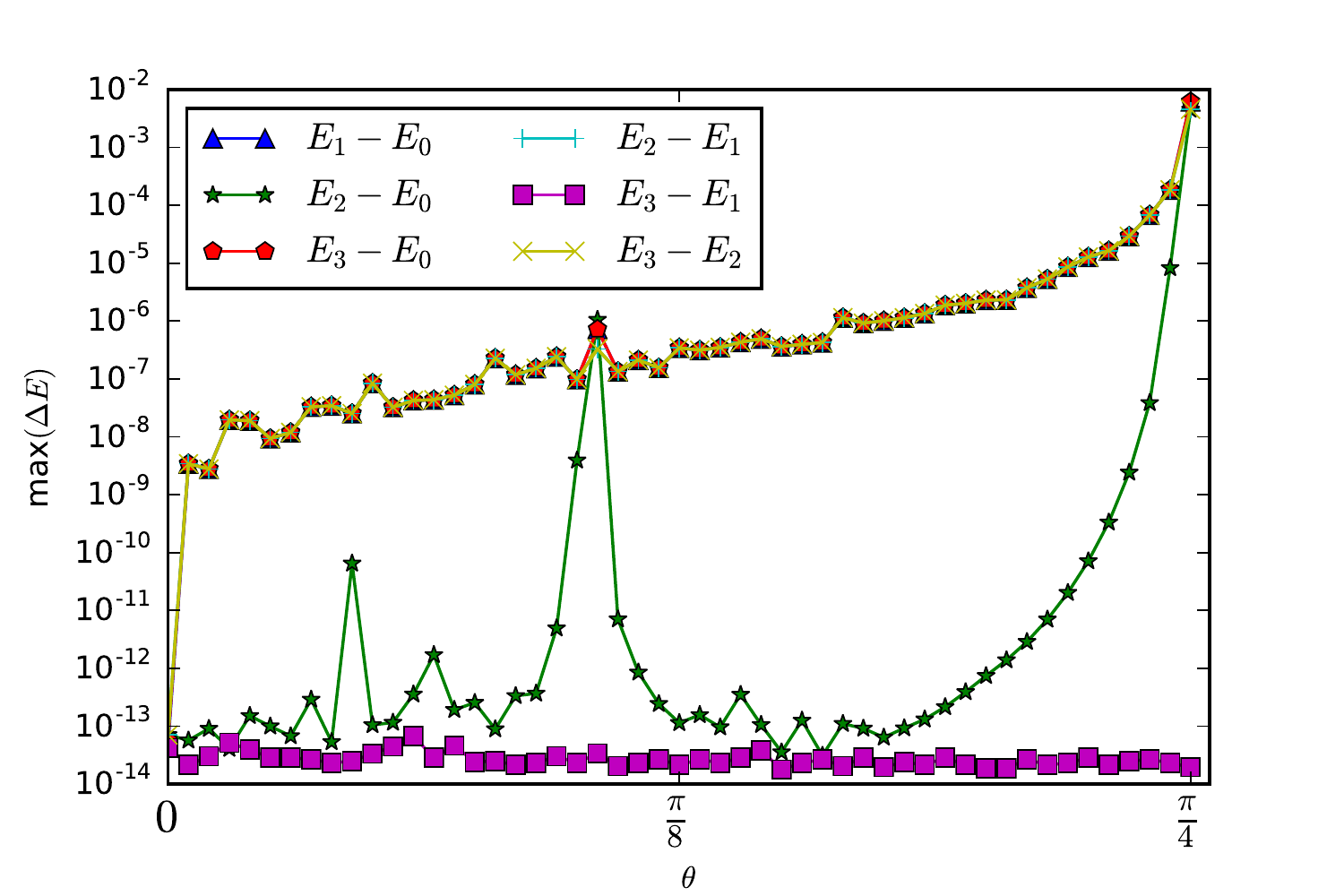}
\caption{Maximal energy splitting between $q$-sectors at $N=4$, as a function of $\theta$ (for $\phi=0$). 
The maximum is taken over the entire spectrum (obtained by exact diagonalisation), for an $L=8$ chain with $f=0.01$} 
\label{fig:N4splitting}
\end{figure}

\begin{figure}
\includegraphics[width=0.45\textwidth]{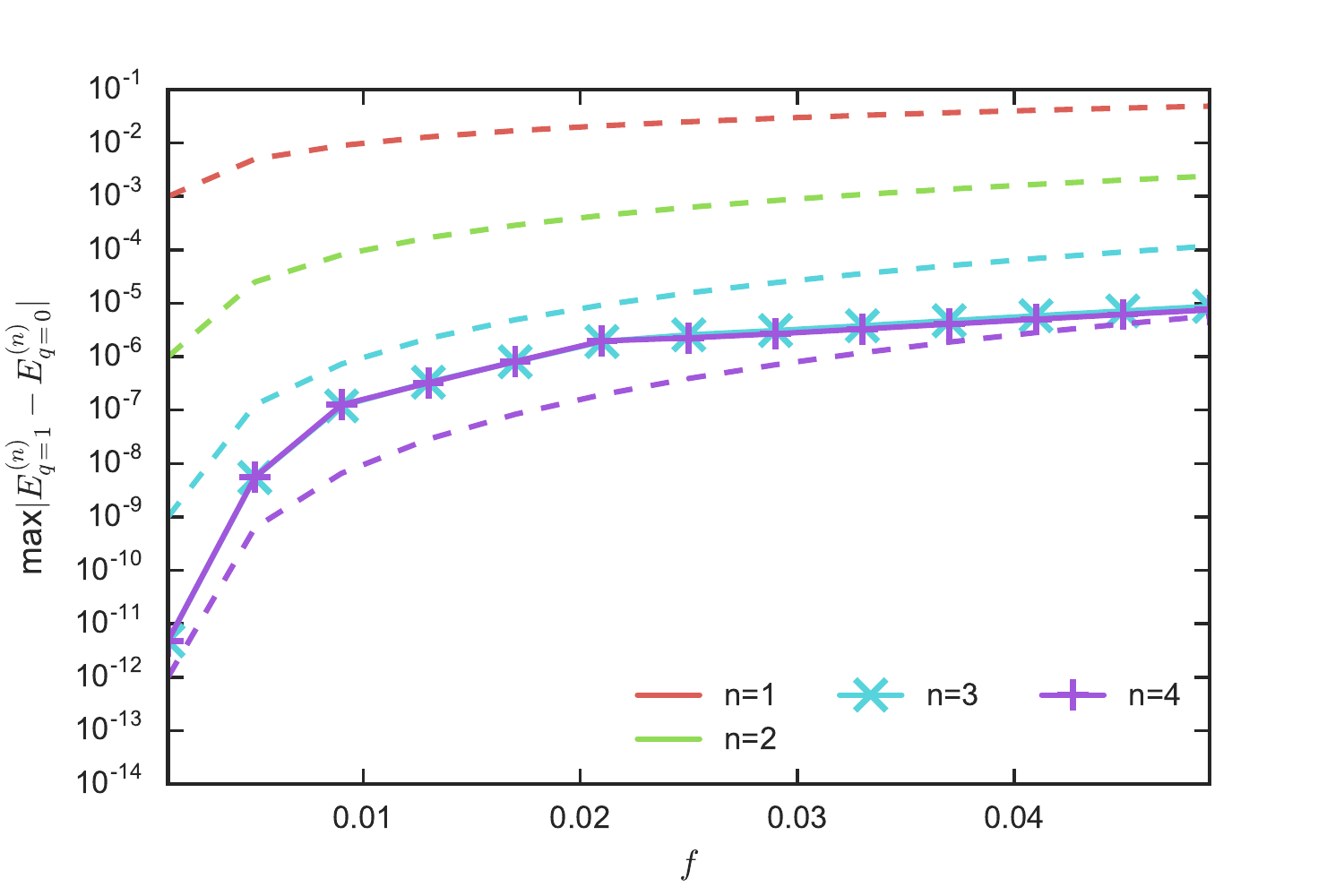}
\caption{The maximum differences between $q=1$ and $q=0$ sectors of the
estimates from $n$th order degenerate perturbation theory for a range of values
of the perturbing parameter $f$. This is for the eighth and ninth excited bands
for a chain of length $L=7$ with $N=4$ and with chiral parameter
$\theta=0.2$. The dashed lines show $f^n$ and act as a
guide. The degenerate space from which we perturb here has dimension 90.} 
\label{fig:PT_N4_bands_8_9}
\end{figure}

As $N=4$ is composite, there are bands which are everywhere degenerate and this
should allow them to be split in a $q$-dependent fashion for generic $\theta$.
The special points above are an exception to this, but once we move away from
these we expect to rapidly loose the strong zero-mode protection, see
section~\ref{sec:unperturbed_system}. This is clear from figure
\ref{fig:N4splitting}, where we show the maximal energy splitting between
$q$-sectors for all energy levels as a function of $\theta$, for a small finite
system. Away from the $\theta=0$ there is generically splitting between all
$q$-sectors which differ by $1$ or $3$. Figure \ref{fig:PT_N4_bands_8_9} shows
the splitting as a function of $f$ at each order of perturbation theory for two
such everywhere degenerate bands of a seven site system. This figure clearly
demonstrates that there are third order processes which split the
degeneracy between $q$-sectors and that this splitting does not disappear as
higher order terms are added.

For $q$-sectors which differ by $2$, there is no splitting observed at generic
$\theta$. The absence of splitting between the $q=1$ and $q=3$ sectors is
explained by the dihedral symmetry discussed in section \ref{sec:dihedral}
(which is present since we set $\phi=0$). However between the $q=0$ and $q=2$ sectors
there is splitting around the most prominent resonance points
(cf.~Fig.~\ref{fig:N4L8Map}), but behaviour consistent with exact degeneracy away
from these points. 

The fact that that the sectors with $\Delta q =2$ are not generally split points
to the survival of order two zero-mode operators related to $\alpha_{L}^{2}$ and
$\alpha_{R}^{2}$. To see this note that whenever we have a $\ZZ_{N}$
parafermionic zero-mode such that $N$ is even, then the operator
$\Phi=\Psi^{(N/2)}$ is a zero-mode with $\Phi^2=1$ and $\Phi Q=-Q\Phi$.
Considering the two sides of the chain, we also note that
$\Phi_{L}\Phi_{R}=(-1)^{N/2}\Phi_{R}\Phi_{L}$, so $\Phi$ is a fermionic mode if
$N/2$ is odd. If $N/2$ is even, as is the case here then $\Phi$ is bosonic, but
its existence nevertheless requires order two degeneracy throughout the
spectrum, as it anti-commutes with $Q$. At $\theta=0$ we can immediately write
$\Phi_{L}=(\alpha_{L})^{2}=\alpha_{L}^{u}\alpha_{L}^{d}$ and $\Phi_{R}=$ for any
$f$. Away from $\theta=0$, the parafermionic zero-modes no longer exist but the
mode $\Phi$ may survive. We can attempt to construct it using an iterative
technique similar to that used in Ref.~\onlinecite{Fendley2014} which yields a
power series expansion in the parameter $f/J$. We write
\begin{equation}
\Phi(\theta)=\sum_{p=0}^{\infty} \left(\frac{f}{J}\right)^{p} \Phi^{(p)} (\theta)
\end{equation}
and require that 
\begin{eqnarray}
[\Phi,H]&=& [\Phi^{(0)},H_0]\\
&+&\sum_{m=1}^{\infty} \left(\frac{f}{J}\right)^{m} \left( [\Phi^{(m-1)},V]+[\Phi^{(m)},H_0]\right) =0,  \non 
\end{eqnarray}
where $H_0=H_J/J$ and $V=H_f/f$. 
We can now determine $\Phi^{(n)}$ from $\Phi^{(n-1)}$, determining the entire
series from $\Phi^{(0)}$, by requiring that
$[\Phi^{(m)},H_0]=-[\Phi^{(m-1)},V]$. Taking
$\Phi^{(0)}=\sigma^{z}_{1,u}\sigma^{z}_{1,d}$ (which is clearly correct at
$\theta=0$), we obtain, to first order in $\frac{f}{J}$,
\begin{eqnarray}
\Phi&=&
\sigma^{z}_{1,u}\sigma^{z}_{1,d} \\
&+&
\frac{f}{J}\frac{\sigma^{x}_{1,u}}{\cos(2\theta)}
\left(\cos(\theta)\sigma^{z}_{d,1}\sigma^{z}_{u,2} - \sin(\theta)\sigma^{z}_{d,1}\sigma^{z}_{d,2}\right)  \non \\
&+&
\frac{f}{J}\frac{\sigma^{x}_{1,d}}{\cos(2\theta)}
\left(\cos(\theta)\sigma^{z}_{u,1}\sigma^{z}_{d,2} + \sin(\theta)\sigma^{z}_{u,1}\sigma^{z}_{u,2}\right)
 \non
\end{eqnarray}
The number of terms on the right hand side grows fast with the order of
approximation. For example $\Phi^{(2)}$ involves terms with products of the
operators $\sigma^{x}_{u/d,1/2}$ and $\sigma^{z}_{u/d,1/2/3}$ and we spare the
reader the explicit expression. We do not know whether the operator $\Phi$ is
generically normalizable. However, it is clear that even at first order, there
are singularities in the expansion. We note that the $\cos(2\theta)$ in the
numerator of $\Phi^{(1)}$ becomes zero precisely at the resonance point
$\theta=\pi/4$ where there is indeed $q$-dependent splitting of the degeneracy
at first order in $f$ (this is prominently visible in
Fig.\ref{fig:N4splitting}). We expect that, similarly to the parafermionic mode
at $N=3$, the zero-mode $\Phi$ at $N=4$ has an expansion to arbitrarily high
orders at generic $\theta$, which may work well in finite systems but which will
have a radius of convergence equal to zero at all $\theta$ in the 
$L\rightarrow \infty$ limit.

\section{Conclusions and outlook}
\label{sec:conclusions}

In this work we looked into the presence of strong zero-modes in $\ZZ_N$
parafermionic chain models. We did this through a detailed study of topological
degeneracies in the entire energy spectra of these models, a necessary condition
for strong zero-modes. When $N>2$ is prime, and in particular when $N=3$, the
introduction of a chiral parameter $\theta$ can lead to regions where the
necessary degeneracy exists is of order $f^L$. However, in the thermodynamic limit,
resonance points where bands of the unperturbed model cross become dense on the
$\theta$-axis and break the degeneracy at a lower order, at least in states
which are nearby in energy. However not all resonance points result in splitting
of the energy and we find so-called 'anti-resonance' points where no such
splitting occurs. These are found where bands with the same total domain-wall
angle cross. A prominent example of this for $N=3$ is at $\theta=\frac{\pi}{6}$.
While there are many other 'anti-resonance' points for $N=3$, the point at
$\theta=\frac{\pi}{6}$ is particularly interesting because the energy difference
between bands here remains of order $J$ as $L \to \infty$. This suggests that 
 strong zero-modes could persist at this point. It would be interesting to further
 explore the existence and structure of these modes using iterative methods
 similar to those employed in Refs. \onlinecite{Fendley2012, JackKemp2017} and
 numerical methods similar to those discussed in Ref. \onlinecite{Kells2015b}.
For $N=4$ 'anti-resonance' points are found at the achiral points. Here the
model is exactly solvable and expressions for the parafermionic zero-modes were
derived.

Our approach includes the introduction of a basis of domain-wall states which
are also eigenstates of the topological symmetry $Q$. This enables us to write
the hamiltonian directly in each $q$-sector and to isolate the $q$-dependent
terms on the end of the chain. We used this to show directly that there can be
no first or second order processes splitting the degeneracy between $q$-sectors
at off-resonant points for general $N$, and none at third order for $N=3$. Using
numerical perturbation theory methods we were able to show for accessible prime
$N>2$ (notably $N=3$) that at off-resonant points, there is no splitting between
$q$-sectors on orders less than the chain length. We were also able to numerically
pick out the order at which the degeneracy is split at particular resonance
points, obtaining results consistent with a lower bound for the order of
splitting, given in Eq.~(\ref{eq:orderbound}). We showed that by excluding the
unlinked processes responsible for $q$-dependent splitting in the pertubative
expansions, the perturbative approximations still converge to the exact values.
While it is not clear a priori that this is a valid perturbative expansion, it
nevertheless shows that the unlinked processes cancel to the order of
approximation that we have calculated, and we expect this to continue at higher
orders.

We also showed that when $N$ is not prime, there is qualitatively different
behaviour resulting from pairs of bands which remain degenerate for all
values of $\theta$. This causes energy splitting at fixed order over the entire
$\theta$ axis (bar some exceptional points). Hence there are generically no
strong parafermionic zero-modes when $N$ is not prime. However at special values
of $N$ and $\theta$, strong parafermionic zero-modes do occur. In particular
this is the case at $N=4$ and $\theta=\frac{m\pi}{2}$. These points are special
as the model can be rewritten as a pair of Ising chains and solved exactly for
all $f$ and $J$. The $N=4$ model also has interesting features away from the
special $\theta$ values. For example, despite its lack of parafermionic
zero-modes it does feature a bosonic zero-mode which again has convergence issues
at resonance points, but only at those resonance points which involve pairs of
bands which are not everywhere degenerate.

Another aspect of this work is the analysis of symmetries from section
\ref{sec:dihedral}. When $\theta$ is a multiple of $\frac{\pi}{N}$ and $\phi=0$,
the models have both a conjugation (time reversal) and a spatial parity
symmetry. For general $\theta$, but still at $\phi=0$, these are no longer
symmetries individually but their product is still a symmetry, which does not
commute with $Q$. This gives us the non-Abelian dihedral group as a symmetry.
From the representation theory of this group we can determine that (at
$\phi=0$), the $q$ and $N-q$ sectors are exactly degenerate for each $N$.

A natural direction for further work is to characterize the energy scales below
which degeneracies can persist in the thermodynamic limit as a function of
$\theta$. A look at Fig.~\ref{fig:N3L30respoints} which shows the resonance
points for an $L=30$ chain shows clearly that the lowest energy at which a
resonance point can be found varies considerably with $\theta$. In fact even at
this relatively large size, there are still considerable windows without any
resonance points around $\theta=0$, $\theta=\frac{\pi}{6}$, and a number of other
prominent resonance points. We have shown that the window around the
$\theta=\frac{\pi}{6}$ point closes as $\delta \theta \sim \frac{1}{\sqrt{3} L}$ and that
the order of the processes that can cause splitting at the closest resonance
point increases linearly with $L$. It would be interesting to see if this is a 
particular feature of the $\theta=\frac{\pi}{6}$ point or if similar behaviour is
observed around other anti-resonance points.

Another natural question is how widely the influence of a particular resonance
extends along the $\theta$-axis. In other words, how far from the resonance
point can we expect appreciable $q$-dependent splitting and how does this length
scale on the $\theta$-axis scale with $L$? In Ref.~\onlinecite{Jermyn2014} the
scaling behaviour of the energy splitting for the lowest states of the first excited
band around $\theta=0$ was explored numerically. It was found that the width of
the region of influence around this particular resonance point decays rapidly
with $L$ which suggests that similar behaviour might be observed at other
resonance points.

\bibliography{corr}
\bibliographystyle{h-physrev3}

\appendix

\section{Degenerate perturbation theory formalism}\label{appx:perturbation_theory}
In this appendix we outline some details of the Raleigh-Schr\"{o}dinger
degenerate perturbative expansions that are used in this work. For further
details and full derivations, consult
Refs~\onlinecite{Kato1949,Bloch1958,Bloch1958b,Messiah1962,Lowdin1962}.

We consider hamiltonians of the form $H=H_0 + \lambda V$, where $H_0$ is the
unperturbed part, and $\lambda V$ the perturbation. We focus on an eigenspace of
of $H_0$ with energy $E_0$. The projector onto this eigenspace is written $P_0$
and we write $Q_0=1-P_0$ for its complement. We also define the operator $a =
E_0 - H_0$ which gives the difference between the unperturbed energy of the
chosen eigenspace and the unperturbed energy of the state on which we apply it.

We use the version of the formalism developed by Bloch\cite{Bloch1958}. Here,
the eigenvalues of $H$ which reduce to $E_0$ as $\lambda\rightarrow 0$ are
approximated to order $\lambda^n$ by the eigenvalues of an effective hamiltonian
\[
H^{\text{eff} (n)} = P_0 H_0 P_0 + \sum_{j=0}^{n-1} P_0 \lambda^{j+1} V \mathcal{U}^{(j)} 
\]
Note that this acts non-trivially only on the chosen eigenspace of $H_0$. 
The $U^{(j)}$ are given by $U^{(0)}=P_0$ and for $j>0$,
\[
\mathcal{U}^{(j)} = \sum_{\vec{k}}^{(j)}S^{k_1}VS^{k_2}V...VS^{k_j}VP_0,
\]
where $\sum_{\vec{k}}^{(j)}$ is a sum over all sets of non-negative integers $k_1,
k_2,...,k_j$ satisfying the conditions:
\begin{eqnarray*}
k_1 &+ k_2 + ... + k_p \ge p &\qquad (p=1,2,...,j-1)\\
k_1 &+ k_2 + ... + k_j = j&
\end{eqnarray*}
and the $S^k$ are given by
\begin{equation}
S^k = 
\begin{cases}
-P_0 \qquad &if \quad k =0\\
\frac{Q_0}{a^k}\qquad &if \quad k\ge 1.
\end{cases}
\end{equation}
All these expressions are slightly modified from earlier work by
Kato\cite{Kato1949} who showed that the expansion converges absolutely when
$||\lambda V || < \Delta_0(E_0)/2$, where $\Delta_0(E_0)$ is the distance from
$E_0$ to the nearest eigenvalue $E\neq E_0$ of $H_0$.

The $\mathcal{U}^{(j)}$ can be efficiently calculated using $\mathcal{U}^{(0)} =
P_0$ and the recurrence relation
\begin{equation*}
\mathcal{U}^{(j)} = \frac{Q_0}{a}\left( V \mathcal{U}^{(j-1)} - \sum_{k=1}^{j-1}\mathcal{U}^{(k)}V\mathcal{U}^{(j-k-1)} \right) 
\end{equation*}
To fourth order, the explicit effective hamiltonian is 
\begin{align}
& H^{\text{eff} (4)}  =  P_0 (H_0 + \lambda V)P_0 + \lambda^2 P_0 V \frac{Q_0}{a}VP_0 + \non \\
  & + \lambda^3\left( P_0 V \frac{Q_0}{a}V\frac{Q_0}{a} V P_0 - 
       P_0 V  \frac{Q_0}{a^2}V P_0 V P_0\right) \non \\ 
  & + \lambda^4\left( P_0 V \frac{Q_0}{a}V\frac{Q_0}{a} V \frac{Q_0}{a}V P_0 -
       P_0 V \frac{Q_0}{a^2}V\frac{Q_0}{a} V P_0 V P_0  \right. \non\\ 
      & - \left. P_0 V \frac{Q_0}{a}V\frac{Q_0}{a^2} V P_0 V P_0 +  
       P_0 V \frac{Q_0}{a^3}V P_0 V P_0 V P_0 \right)  \label{eqn:bloch_to_fourth_order}
\end{align}
Up to order $\lambda^{n}$, the eigenvectors of $H^{\text{eff}(n)}$ in the $E_0$
eigenspace of $H_0$ are the projections of the corresponding eigenvectors of $H$
to this eigenspace. While the eigenvalues of $H$ are orthogonal, their
projection onto the $E_0$ eigenspace of $H_0$ need not be. This is reflected by
the fact that $H^{\text{eff}(n)}$ is typically not Hermitian for $n\ge 3$.

\section{Vanishing of third and higher order corrections in off-resonant perturbation theory}
\label{appx:3rdorder}
Here we look at the third and higher order pertubative expansions at
off-resonant points. We first talk about how at orders $n \ge 2$ the choice of
expansion used becomes important. Following this we show a specific example of a
third order processes that leads to $q$-dependent matrix elements when using
Bloch's expansion. Next we show that any $q$-dependent third order processes
supported at non-overlapping positions cancel. Finally we consider the third
order processes at overlapping positions and show that using Soliverez's
expansion these cancel in the specific case mentioned earlier, and using
numerics cancel in general for $N=3$. 

\subsection{Different pertubative expansions}
\label{appx:pt_exps}
An additional complicating issue, which appears in degenerate perturbation
theory, is that there is no canonical choice of perturbation series. In fact
there are multiple perturbation series which converge to the same limit (when
they converge), but which can differ from each other at any finite order by
higher order contributions. For example, in our numerical examinations we use
the expansion by Bloch\cite{Bloch1958} which involves diagonalising effective
hamiltonians which are completely $q$-independent to second order in $f$ but
which do depend on $q$ at order $f^3$ and onward. On the other hand, the
effective hamiltonians in the expansion by Kato\cite{Kato1949} (of which Bloch's
is a modification) actually have $q$-dependent terms already at order $f^2$,
despite the fact that the two expansions converge to the same limit. Another
expansion by Soliverez\cite{Soliverez1969} also has eigenvalues which converge
to the same limit, but has effective hamiltonians which are independent of $q$
to order $f^3$. This means in particular that there is no $q$-dependent energy
splitting up to order $f^3$, despite the fact that the Kato and Bloch effective
hamiltonians already depend on $q$ at this order. Unfortunately the hamiltonians
in Soliverez' expansion also become $q$-dependent from the next order up. While
it may be possible to devise a perturbation scheme tailor made for this problem
which has an explicitly $q$-independent effective hamiltonian up to order
$f^{L-1}$ we will not pursue this here.

\subsection{Specific example}
First of all, let us show explicitly that the effective hamiltonian $H^{\text{eff}(3)}$
has nonzero $q$-dependent matrix elements. 
For small $\theta\neq 0$ we consider the first band above the ground state
consisting of states $\ket{1_x}_q$. Within this band, $H^{\text{eff}(3)}$
connects the states $\ket{1_2}$ and $\ket{1_{L-1}}$, which host domain-walls on
the second and last sites, respectively. The terms in $H^{\text{eff}(3)}$ which connect
these states must involve the operators $\alpha_1^{\phantom{\dagger}}$,
$\alpha^{\dagger}_{1} \alpha_2^{\phantom{\dagger}}$ and $\alpha^{\dagger}_{L-1}$
from $H^{q}_{f}$. These can potentially appear in the third order terms of
$H^{\text{eff}(3)}$ in six possible permutations. Applying each of these permutations
we find, up to multiplication by an overall factor of $\omega^{q} f^3 e^{3
i\phi}$,
\begin{align}
\label{eq:perms}
\ket{1_2}&
\xrightarrow{\alpha_1^{\phantom{\dagger}}}
\ket{(N\!-\!1)_1 1_2}\xrightarrow{\alpha^{\dagger}_{1}\alpha_{2}^{\phantom{\dagger}}}\ket{\emptyset}
\xrightarrow{\alpha^{\dagger}_{L\!-\!1}}\ket{1_{L\!-\!1}}
\non \\
\ket{1_2}&\xrightarrow{\alpha_1^{\phantom{\dagger}}}\ket{(N\!-\!1)_1 1_2}\xrightarrow{\alpha^{\dagger}_{L\!-\!1}}\ket{(N\!-\!1)_1 1_2 1_{L\!-\!1}}
\xrightarrow{\alpha^{\dagger}_{1}\alpha_{2}^{\phantom{\dagger}}}\ket{1_{L\!-\!1}}
\non\\
\ket{1_2}&\xrightarrow{\alpha^{\dagger}_{1}\alpha_{2}^{\phantom{\dagger}}}\ket{1_1 }\xrightarrow{\alpha_{1}^{\phantom{\dagger}}}\ket{\emptyset}
\xrightarrow{\alpha^{\dagger}_{L\!-\!1}}\ket{1_{L\!-\!1}}
\non\\
\ket{1_2}&\xrightarrow{\alpha^{\dagger}_{1}\alpha_{2}^{\phantom{\dagger}}}\ket{1_1}\xrightarrow{\alpha^{\dagger}_{L\!-\!1}}\ket{1_1 1_{L\!-\!1}}
\xrightarrow{\alpha_{1}^{\phantom{\dagger}}}\ket{1_{L\!-\!1}}
\non\\
\ket{1_2}&\xrightarrow{\alpha^{\dagger}_{L\!-\!1}}\ket{1_2 1_{L\!-\!1}}\xrightarrow{\alpha_{1}^{\phantom{\dagger}}}\ket{(N\!-\!1)_1 1_2 1_{L\!-\!1}}
\xrightarrow{\alpha^{\dagger}_{1}\alpha_{2}^{\phantom{\dagger}}}\ket{1_{L\!-\!1}}
\non\\
\ket{1_2}&\xrightarrow{\alpha^{\dagger}_{L\!-\!1}}\ket{1_2 1_{L\!-\!1}}\xrightarrow{\alpha^{\dagger}_{1}\alpha_{2}^{\phantom{\dagger}}}\ket{1_1 1_{L\!-\!1}}
\xrightarrow{\alpha^{\phantom{\dagger}}_{1}}\ket{1_{L\!-\!1}}
\end{align}
We now see that the processes on the third and fourth lines here do not actually
lead to terms in $H^{\text{eff}(3)}$. This is because $\ket{1_1}$ is in the same band
$b$ as the the initial and final states and is thus projected out by the
$Q_{0}$ operator. In terms of the domain-wall
energies $\epsilon_d$ in (\ref{eqn:DWH0Energy}) we then find
\begin{align*}
\frac{\bra{1_2}H^{\text{eff}(3)}\ket{1_{L\!-\!1}}}{\omega^q f^3 e^{3i\phi}}=
& \frac{1}{\epsilon_0\!-\!\epsilon_{N\!-\!1}}\left[ \frac{1}{\epsilon_1\!-\!\epsilon_0} + \frac{1}{2\epsilon_0\!-\!\epsilon_1\!-\!\epsilon_{ N\!-\!1}} \right] \\
+&\frac{1}{\epsilon_0\!-\!\epsilon_1}\left[ \frac{1}{2\epsilon_0\!-\!\epsilon_1\!-\!\epsilon_{N\!-\!1}} + \frac{1}{\epsilon_0\!-\!\epsilon_1}  \right] 
\\
  =& \frac{1}{(\epsilon_0 \!-\! \epsilon_1)^2}
\end{align*}
which shows that $H^{\text{eff}(3)}$ has at least one element that depends on $q$. We
will comment on more general analytical calculation of $q$-dependent matrix
elements of effective hamiltonians in the next subsection.

\subsection{General arguments for the absence of splitting}

All of the perturbative schemes mentioned in section \ref{appx:pt_exps} involve the
evaluation of the matrix elements of a characteristic set of operators at each
order. For example, at third order, the following operators play a role
\begin{eqnarray}
O_{11}&=&P_{0} H^{q}_{f} \frac{Q_{0}}{E_{0}-H_0}H^{q}_{f}\frac{Q_{0}}{E_{0}-H_0} H^{q}_{f} P_{0} 
\nonumber \\
O_{20}&=&P_{0} H^{q}_{f} \frac{Q_{0}}{(E_{0}-H_0)^2} H^{q}_{f} P_{0} H^{q}_{f} P_{0}
\nonumber \\
O_{02}&=&P_{0} H^{q}_{f} P_{0} H^{q}_{f} \frac{Q_{0}}{(E_{0}-H_0)^2}  H^{q}_{f} P_{0}
\end{eqnarray}
At order $n$, we have operators $O_{p_1...p_{n-1}}$ containing $n-1$ factors
$(E_0-H_0)^{-p_{i}}$ with $p_{i}\ge 0$ and $\sum{p_i}=n-1$. The matrix elements
of these operators are built from the various terms in $H^{q}_{f}$. Each
occurrence of $H^{q}_{f}$ may be replaced by either $\alpha_1$, $\alpha_{L-1}$
or $\alpha^{\dagger}_{x}\alpha_{x+1}$ or their complex conjugates, with the
appropriate prefactor from Eq.~(\ref{eqn:Hf}). We can ignore the prefactors for
now and note that the $\alpha_{x}$ have matrix elements which can only take the
values $0$ and $1$.

\subsubsection{Non-overlaping positions}
If we take three terms from $H^{q}_{f}$ which are supported
at non-overlapping positions (for example $\alpha_1$, $\alpha^{\dagger}_{L-1}$
and $\alpha^{\dagger}_{x}\alpha_{x+1}$ with $2\le x \le L-3$) then it is easy to
calculate the corresponding contributions to the matrix elements of all three
operators above. Applying each of the three operators causes a change in the
energy, let's call these energy differences $a$, $b$ and $c$, so for instance we
may have the situation that applying $\alpha_1$ to a state with energy $E$ gives
a state with energy $E+a$, subsequently applying
$\alpha^{\dagger}_{x}\alpha_{x+1}$ to the state will give an energy $E+a+b$ and
finally applying $\alpha^{\dagger}_{L-1}$ gives energy $E+a+b+c$. We can only
get a nonzero matrix element if the final state is in the same band as the
initial state, which means we must require $a+b+c=0$. Assuming that $a\neq 0$
and $a+b \neq 0$ this set of operators will only give a contribution to $O_{11}$
and this will be equal to $\frac{1}{a(a+b)}$ up to the appropriate factors from
Eq.~(\ref{eqn:Hf}). However the triple of operators corresponding to the energy
changes by $a$, $b$ and $c$ can occur in $6$ different orders. If $a$,$b$ and
$c$ (and hence also $a+b$, $a+c$ and $b+c$) are all nonzero, there are $6$
corresponding contributions to any matrix element of $O_{11}$. These all
cancel, since
\begin{eqnarray}
~&\frac{1}{a(a+b)}+\frac{1}{a(a+c)}+\frac{1}{b(a+b)}& \nonumber\\
+&\frac{1}{b(b+c)}+\frac{1}{c(a+c)}+\frac{1}{c(b+c)}&
=\frac{a+b+c}{abc}
\end{eqnarray}
which vanishes, since we know that $a+b+c=0$. Similar identities make many
contributions to the matrix elements of the operators $O_{11...1}$ which occur
at higher orders vanish.

If one of the energy differences $a$, $b$ or $c$ equals zero, the contribution
to the corresponding matrix element of $O_{11}$ no longer vanishes, but can in
principle be cancelled by contributions to the matrix elements of $O_{02}$ or
$O_{20}$. For example if $a=0$ then necessarily also $b+c=0$ and there are only
two ways to produce a non-vanishing contribution to $O_{11}$, by adding energy
according to the following steps,
\begin{eqnarray*}
E\mathrel{\mathop{\rightarrow}^{\mathrm{+b}}}  E+b \mathrel{\mathop{\rightarrow}^{\mathrm{+a}}} E+b \mathrel{\mathop{\rightarrow}^{\mathrm{+c}}} E\\
E\mathrel{\mathop{\rightarrow}^{\mathrm{+c}}}  E-b \mathrel{\mathop{\rightarrow}^{\mathrm{+a}}} E-b \mathrel{\mathop{\rightarrow}^{\mathrm{+b}}} E.
\end{eqnarray*}
These give a total contribution proportional to $\frac{2}{b^2}$ (or equally
$\frac{2}{c^2}$). The other orders of adding the energy now yield contributions
to $O_{02}$ and $O_{20}$. We find that these contributions are also proportional
to $\frac{2}{b^2}$ for $O_{02}$ (from $a+b+c$ and $a+c+b$) and again
proportional to $\frac{2}{b^2}$ for $O_{20}$ (from $b+c+a$ and $c+b+a$), all
with the same proportionality constants from Eq.~(\ref{eqn:Hf}). Similarly, if
$b=0$ or $c=0$, we find that all operators get contributions proportional to
$\frac{2}{a^2}$. It is clear that the contributions from $O_{11}$, $O_{02}$ and
$O_{20}$ can all cancel each other, depending on how these operators appear in
the effective hamiltonian. The Bloch hamiltonian $H^{\text{eff}(3)}$ contains the
combination $O_{11}-O_{20}$ and does not contain $O_{02}$. We see that the
contributions we have been considering here all vanish. The same is true for
Soliverez' third order hamiltonian,
\[
H^{Sol(3)}_{q}
=\frac{1}{2}(H^{\text{eff}(3)}_{q}+(H^{\text{eff}(3)}_{q})^{\dagger})=
O_{11}-\frac{1}{2}O_{20}-\frac{1}{2}O_{02}
\]
Again, the contributions from the $O_{p_1 p_2}$ cancel.

\subsubsection{Overlaping positions}
We see that, in order to have non-vanishing contributions, we must take
operators from $H^{q}_{f}$ which have overlapping support, that is, at least two
of the operators form a ``chain''. An examples is the triple $\alpha_{1}$,
$\alpha^{\dagger}_{1}\alpha^{\phantom{\dagger}}_{2}$, $\alpha_{L}$ which we
analysed in section~\ref{appx:3rdorder}. In such cases the energy differences
which appear depend on the order of application of the operators and we showed
already that $\bra{1_2} H^{\text{eff}(3)}
\ket{1_{L-1}}=(\epsilon_{0}-\epsilon_{1})^{-2}\neq 0$. We also noted during the
calculation that in this case $\bra{1_2} O_{20} \ket{1_{L-1}}=0$. We can also
calculate $\bra{1_2} O_{02} \ket{1_{L-1}}$, which appears in Soliverez'
hamiltonian. We see that the third and fourth permutations shown in
Eq.~(\ref{eq:perms}) give
\begin{equation*}
\bra{1_2} O_{02} \ket{1_{L-1}}= \frac{1}{(\epsilon_1-\epsilon_0)^2} + \frac{1}{(\epsilon_0-\epsilon_1)^2} =\frac{2}{(\epsilon_1-\epsilon_0)^2}
\end{equation*}
The contribution from $O_{02}$ to Soliverez' hamiltonian thus precisely cancels
the contribution from $O_{11}$ and this potentially $q$-dependent matrix element
is actually zero. One may similarly show that all further potentially
$q$-dependent matrix elements of Soliverez' third order hamiltonian are zero, in
all bands (assuming $L\ge 3$). This shows in particular that there is no
$Q$-dependent energy splitting in any band at order $f^3$, as long as $\theta$
is not at a resonance point.

We should note that all results we have presented about the absence of
$Q$-dependent energy splitting at first, second and third order in $f$ are still
valid if the coefficients $f$ and $J$ are allowed to vary along the chain
(assuming they remain bounded away from zero). This is easy to see in
retrospect. All results on vanishing were obtained by showing cancellation of
contributions from sets of operators taken from of $H^{q}_{f}$ acting in
different orders but at fixed positions along the chain. The cancellations are
never dependent on any relationship between coupling constants at different
sites or links of the chain.
  
\end{document}